\documentclass[11pt,a4paper]{article}
\usepackage{bbm}
\usepackage{feynmf}
\usepackage{jheppub}
\usepackage{mathrsfs}
\usepackage{mathtools}

\allowdisplaybreaks

\let\a=\alpha
\let\b=\beta
\let\g=\gamma
\let\d=\delta
\let\eps=\epsilon
\let\k=\kappa
\let\l=\lambda
\let\m=\mu
\let\n=\nu
\let\p=\phi
\let\r=\rho
\let\s=\sigma
\let\t=\tau
\let\c=\chi
\let\O=\Omega
\let\x=\xi
\let\de=\partial
\let\w=\wedge

\newcommand{\im}{\text{i}}
\newcommand{\un}{\mathbbm{1}}
\newcommand{\gr}[1]{\mathrm{#1}}
\newcommand{\dd}{\text{d}}
\newcommand{\D}{\text{D}}
\newcommand{\Da}{\text{D}}
\newcommand{\DD}{\mathbbm{D}}
\newcommand{\abs}[1]{\lvert#1\rvert}
\newcommand{\K}{\mathcal{K}}
\renewcommand{\L}{\mathscr{L}}
\newcommand{\E}{\mathcal{E}}
\newcommand{\X}{\mathcal{X}}
\newcommand{\mc}{\omega}
\newcommand{\Piinv}{\rotatebox[origin=c]{180}{$\Pi$}}
\DeclareMathOperator{\ch}{ch}
\DeclareMathOperator{\sh}{sh}
\DeclareMathOperator{\arctanh}{arctanh}
\DeclareMathOperator{\tr}{tr}

\newcommand{\rev}[1]{#1}

\newcommand{\feynSSG}[4]{\parbox{20mm}{\begin{fmfgraph*}(20,20)
\fmfleft{l}
\fmfrightn{r}{2}
\fmf{plain,label=$#4$,l.si=right}{r2,v}
\fmf{plain}{v,l}
\fmf{photon}{r1,v}
\fmfv{label=$#1$}{l}
\fmfv{label=$#2$}{r2}
\fmfv{label=$#3$}{r1}
\fmfdot{v}
\end{fmfgraph*}}}

\newcommand{\feynSSSS}[4]{\parbox{20mm}{\begin{fmfgraph*}(20,20)
\fmfleft{l}
\fmfrightn{r}{3}
\fmf{plain}{r2,v,l}
\fmffreeze
\fmf{plain}{r1,v}
\fmf{plain}{r3,v}
\fmfv{label=$#1$}{l}
\fmfv{label=$#3$}{r2}
\fmfv{label=$#2$}{r3}
\fmfv{label=$#4$}{r1}
\fmfdot{v}
\end{fmfgraph*}}}

\newcommand{\feynGSS}[5]{\parbox{20mm}{\begin{fmfgraph*}(20,20)
\fmfleft{l}
\fmfrightn{r}{2}
\fmf{photon}{v,l}
\fmf{plain,label=$#4$,l.si=right}{r2,v}
\fmf{plain,label=$#5$,l.si=left}{r1,v}
\fmfv{label=$#1$}{l}
\fmfv{label=$#2$}{r2}
\fmfv{label=$#3$}{r1}
\fmfdot{v}
\end{fmfgraph*}}}

\title{Geometric duality between effective field theories I: scattering amplitudes}

\author[a]{Tom\'a\v{s} Brauner,}
\author[b]{Yang Li,}
\author[b]{Diederik Roest}
\author[b]{and Tianzhi Wang}

\affiliation[a]{Department of Mathematics and Physics, University of Stavanger,\\
4036 Stavanger, Norway}
\affiliation[b]{Van Swinderen Institute for Particle Physics and Gravity, University of Groningen,\\
9747 Groningen, The Netherlands}

\emailAdd{tomas.brauner@uis.no}
\emailAdd{y.l.li@rug.nl}
\emailAdd{d.roest@rug.nl}
\emailAdd{tian.wang@rug.nl}

\abstract{We propose a novel type of duality that connects a sequence of well-known theories with even-multiplicity scalar amplitudes: it relates the Yang-Mills theory coupled to a specific scalar matter sector to the nonlinear sigma model on a symmetric coset space, the (multiflavor) Dirac-Born-Infeld theory, and the special Galileon theory. The duality is manifested with the help of a covariant formulation of the classical equations of motion that features a contact quartic scalar self-coupling combined with propagation on a dynamical background of elementary or composite gauge fields. This is augmented with a set of constitutive relations that reflect the intrinsic or extrinsic geometry of the target space of the theory. The universality of the underlying geometric structure allows for an unambiguous mapping between different theories.}

\keywords{Effective Field Theories, Scattering Amplitudes, Spontaneous Symmetry Breaking}

\arxivnumber{2505.15199}

\begin{document}

\begin{fmffile}{feynman}

\maketitle


\section{Introduction}
\label{sec:intro}

The last couple of decades have witnessed a revolution in our understanding of perturbative quantum field theory. The traditional framework based on Feynman diagrams is plagued with ambiguities due to field redefinitions, and with the combinatorial explosion of higher orders of perturbation theory. Modern on-shell methods for computation of scattering amplitudes (see ref.~\cite{Elvang2015} for an introduction) kill two birds with one stone. On the one hand, working solely with on-shell quantities makes physical predictions manifestly insensitive to the choice of field parameterization by virtue of the Lehmann-Symanzik-Zimmermann (LSZ) reduction. On the other hand, the recursive bottom-up construction of on-shell scattering amplitudes operates at the level of entire amplitudes, thus bypassing the need to relate to individual perturbative Feynman diagrams.

Perhaps even more importantly, changing the perspective on quantum field theory has brought to light new features and structures, entirely invisible to traditional perturbation theory. There is now a growing catalog of quantum field theories where scattering amplitudes of all multiplicities (at least at tree-level) are completely determined by a finite, often small set of data, combined with the fundamental properties of Lorentz invariance, unitarity and analyticity. In gauge theories such as Yang-Mills theory~\cite{Britto2005,Britto2005a} or perturbative gravity~\cite{Bedford2005,Cachazo2005}, a set of seed amplitudes is sufficient provided the scaling properties of all-multiplicity scattering amplitudes in the ultraviolet limit are known. In effective field theories (EFTs) of massless scalars --- Nambu-Goldstone (NG) bosons --- the seed amplitudes have to be combined with additional input on the scaling of scattering amplitudes in the infrared limit~\cite{Kampf2013a,Cheung2016a}.

One of the most astonishing results that the modern scattering amplitude program has produced is the discovery that amplitudes, and in some cases even certain exact nonlinear solutions, of two disparate theories may be intimately related to each other. This phenomenon, colloquially referred to as the \emph{double copy}, has now been observed in a large variety of contexts; see refs.~\cite{Bern2019a,Bern2022,Adamo2022} for recent reviews. \rev{A common approach to the double copy is based on a representation of scattering amplitudes in terms of diagrams constructed using a single type of a fundamental trivalent interaction vertex, where the contribution of every diagram contains a product of two (theory-specific) numerators.} This is known as the Bern-Carrasco-Johansson (BCJ) representation of the amplitude~\cite{Bern:2010ue}.\footnote{\rev{An alternative representation is due to Cachazo, He and Yuan (see e.g.~refs.~\cite{Cachazo:2013gna,Cachazo:2013iea,Cachazo2014a}), where tree-level scattering amplitudes instead are given by contour integrals localized by a set of scattering equations on a punctured sphere, and where the integrand contains the product of two (theory-specific) numerators.}} Similar purely cubic representations of scattering amplitudes in specific theories have been discovered more recently using the classical equation of motion (EoM)~\cite{Monteiro2011,Cheung2021} and combinatorial methods~\cite{Arkani-Hamed:2023mvg,ArkaniHamed2024}.

Reducing the set of basic building blocks to a single cubic vertex is obviously strongly preferable over traditional perturbation theory which often, e.g.~in perturbative gravity or EFTs such as the nonlinear sigma model (NLSM), includes an infinite tower of interaction vertices with increasing complexity. However, there are numerous quantum field theories in which possible scattering processes are restricted to an \emph{even} number of particles, typically as a consequence of global symmetry and the ensuing conservation laws. \rev{A first, well-known example is the NLSM itself; see ref.~\cite{Kampf2013b} for a discussion of its (tree-level) amplitudes and refs.~\cite{Bartsch2024a,ArkaniHamed2024,Arkani-Hamed:2024yvu} for more recent developments. Other notable theories where only even numbers of particles scatter include the} EFTs known to feature exceptional scaling of amplitudes in the soft limit for a single particle:\footnote{In fact, the soft limits can be seen as \emph{defining} properties of these EFTs: all amplitudes can be on-shell constructed by imposing the corresponding specific soft degrees~\cite{Cheung2015a}.} the Dirac-Born-Infeld (DBI) theory, and the special Galileon (sGal) theory~\cite{Cheung2017a}. In these theories, one can always choose a field parameterization such that the ordinary perturbation theory includes only interaction vertices with even valency. While the interactions keep nontrivial Lorentz structure due to the presence of derivatives, the absence of vertices with odd valency greatly reduces the computational cost.

With this in mind, we develop in the present paper a new representation for tree-level scattering amplitudes in the NLSM on symmetric coset spaces, multiflavor DBI and sGal theories, whose fundamental building block in all cases is a \emph{contact} (i.e.~nonderivative) \emph{quartic} vertex. In order to stress the similarity between the different theories, we present their main features schematically in table~\ref{tab:bigpicture}. The baseline used to compare the different theories is provided by the Yang-Mills-scalar (YMS) theory, that is the Yang-Mills theory coupled to scalar matter with specific quartic self-interaction.\footnote{The YMS theory plays the same role for our setup as the so-called biadjoint scalar theory plays in the BCJ formulation of scattering amplitudes as well as the cubic representations based on the classical EoM~\cite{Monteiro2011,Cheung2021} and pure combinatorics~\cite{Arkani-Hamed:2023lbd,Arkani-Hamed:2023mvg}.}  We will show that one can phrase the different EFTs --- whose fundamental degrees of freedom are in all cases a set of scalar fields --- in a fully analogous fashion, provided they are formulated in terms of certain composite ``currents,'' which are symmetric tensors under the Lorentz group. The rank of the tensor corresponds to the soft scaling degree of the scattering amplitudes of the given theory. The ``primary'' form of the EoM is in each case a second-order partial differential equation for the fundamental scalar fields. When rewritten in terms of the tensor current, the order of the EoM is two minus the rank of the current. The structural analogy between the different theories shows most manifestly in a set of descendant EoMs for the tensor currents, all of second order, highlighted in boxes in table~\ref{tab:bigpicture}. 

\begin{table}
\begin{center}
\makebox[\textwidth][c]{\begin{tabular}{c||c|c|c|c}
Theory & YMS & NLSM & DBI & sGal\\
\hline\hline 
Soft degree & $0$ & $1$ & $2$ & $3$ \\
Symmetric tensor & $\p$ & $\mc_\m$ & $\K_{\m\n}$ & $\O_{\m\n\l}$ \\
\hline
Primary EoM & $\boxed{\Da^2\p+\p^3\sim0}$ & $\Da\cdot\mc\sim0$ & $\tr\K\sim0$ & $\nabla^{-1}\cdot\tr\O\sim0$ \\
$1^\text{st}$ level descendant & --- & $\boxed{\Da^2\mc+\mc^3\sim0}$ & $\DD\cdot\K\sim0$ & $\tr\O\sim0$ \\
$2^\text{nd}$ level descendant & --- & --- & $\boxed{\DD^2\K+\K^3\sim0}$ & $\nabla\cdot\O\sim0$ \\
$3^\text{rd}$ level descendant & --- & --- & --- & $\boxed{\nabla^2\O+\O^3\sim0} $\\
\end{tabular}}
\end{center}
\caption{\emph{Schematic outline of the structure of the theories considered in this paper, emphasizing how the different EFTs with positive soft degree adhere to the structure of the YMS theory. The fundamental degrees of freedom of each of the theories are a set of Lorentz scalars subject to a primary EoM. However, we reformulate our EFTs in terms of a tensor that is rank-1 for the NLSM on symmetric coset spaces, rank-2 for the (multiflavor) DBI theory, and rank-3 for the sGal theory. The associated EoMs for these tensors, framed in a box,  are descendants of the primary EoM. The structural equivalence of the theories is our main result, and is substantiated in section~\ref{sec:flavorkinematics}. Upon matching properly the matter sectors and the gauge sectors (not included in this table), we find two sequences of theories related by duality: one including the YMS theory and the NLSM, and the other the NLSM coupled to gravity and the DBI and sGal theories. The notation used in the table conforms to the conventions used in sections~\ref{sec:YMS} to~\ref{sec:Galileon}, and is explained in detail therein.}}
\label{tab:bigpicture}
\end{table}

As is clear from the covariant derivatives in the EoMs in table~\ref{tab:bigpicture}, the matter sectors therein have to be augmented by appropriate gauge sectors. For the baseline YMS theory, this is simply the Yang-Mills part of the theory. Instead, for the different EFTs, it consists of specific \emph{composite} connections for gauge and/or gravitational sectors. Using solely geometric building blocks guarantees manifest covariance of the EoMs under all the (nonlinear) symmetries of the different theories. 

Simultaneous matching of the dynamics of the matter and gauge sectors across the different theories imposes nontrivial constraints, which lead to two separate sequences of theories related by duality. The first sequence relates the YMS theory to the NLSM, while the second one maps the NLSM on a specific coset space, coupled to gravity (which we for the sake of brevity refer to as NLSM\textsubscript{g}), to the multiflavor DBI theory and the sGal theory. Our results therefore offer a simple explanation of the flavor-kinematics duality between these two sets of theories, observed at the BCJ level in ref.~\cite{Neeling2022}. 

The plan of the paper is as follows. In the following sections, we review the relevant details of all the different theories we consider: the YMS theory in section~\ref{sec:YMS}, the NLSM on symmetric coset spaces in section~\ref{sec:NLSM}, the multiflavor DBI theory in section~\ref{sec:DBI}, and the sGal theory in section~\ref{sec:Galileon}. We make the four sections intentionally independent of each other in order to make the content useful also for readers who are only interested in some of the theories. A synthesis of all the material is offered in section~\ref{sec:flavorkinematics}, where we elaborate on the two duality sequences, YMS-NLSM and NLSM\textsubscript{g}-DBI-sGal. In section~\ref{sec:conclusions}, we conclude and offer an outlook for future work. \rev{Finally, some technical details on sample computations within the different theories, and the relations between them, are provided in appendix~\ref{app:4pt_vertices}.}


\section{Yang-Mills-scalar theory}
\label{sec:YMS}

\subsection{Algebraic structure}

We start with a pure gauge theory with a compact gauge group $H_\text{color}$. We will use the notation $Q_\a$ for the matrix generators of $H_\text{color}$, and $A_\m\equiv A^\a_\m Q_\a$ for the gauge field. The corresponding matrix-valued field strength is
\begin{equation}
F_{\m\n}\equiv\de_\m A_\n-\de_\n A_\m-\im[A_\m,A_\n]\quad\text{or}\quad
F^\a_{\m\n}\equiv\de_\m A^\a_\n-\de_\n A^\a_\m+f^\a_{\b\g}A^\b_\m A^\g_\n,
\end{equation}
where the structure constants of the gauge group are defined through $[Q_\a,Q_\b]=\im f^\g_{\a\b}Q_\g$. 

We now couple the pure gauge theory to a set of matrix scalar fields $\Phi^I\equiv\phi^{Ia}Q_a$, where the matrices $Q_a$ are subject to two conditions:
\begin{itemize}
\item The set $Q_a$ spans a representation of $H_\text{color}$ via adjoint action, i.e.~$[Q_\a,Q_a]=\im f^b_{\a a}Q_b$.
\item The commutator of a pair of $Q_a$ lies in the Lie algebra of $H_\text{color}$, i.e.~$[Q_a,Q_b]=\im f^\a_{ab}Q_\a$. 
\end{itemize}
The sets of constants $f^b_{\a a}$ and $f^\a_{ab}$ are subject to consistency conditions stemming from the Jacobi identity for nested commutators. However, we do not need to spell these conditions out explicitly. Likewise, we do not need to make any further assumptions about the precise relationship between the sets of matrices $Q_\a$ and $Q_a$ at this stage. Finally, we assume that the matrix fields $\Phi^I$ carry, by virtue of their ``flavor'' index $I$, a representation of a global symmetry group $G_\text{flavor}$.

For future reference, let us mention explicitly two special cases of particularly high relevance. One possible, and rather obvious, choice of algebraic structure is to identify the Greek and lowercase Latin indices, and likewise the matrices $Q_\a$ and $Q_a$. In this case, the scalar fields $\p^{Ia}$ transform in the adjoint representation of $H_\text{color}$, and the corresponding version of the YMS theory is known to descend from pure Yang-Mills theory via dimensional reduction~\cite{Cachazo2015}. The second, more general possibility is to treat $Q_\a$ and $Q_a$ as (mutually) linearly independent sets of matrices that together comprise the Lie algebra of a larger Lie group, $G$. The matrices $Q_\a$ generate a subgroup, $H\simeq H_\text{color}$, of $G$. This algebraic structure can be naturally interpreted in terms of spontaneous symmetry breaking: $G$ is the complete symmetry group, whereas $H$ is the unbroken subgroup. The requirement that $[Q_a,Q_b]$ lies in the Lie algebra of $H$ means that the coset space $G/H$ is symmetric.


\subsection{Equations of motion}

The next step is to construct a Lagrangian. To that end, we need to be able to pair indices on the fields in a way that respects the symmetry. We will need an invariant positive-definite symmetric bilinear form (metric) $g_{\a\b}$ on the Lie algebra of $H_\text{color}$. This can be taken for instance as the Cartan-Killing form. In addition, we assume that the representation of $H_\text{color}$ carried by $\phi^{Ia}$ possesses an invariant metric, given by $g_{ab}$, and likewise for $G_\text{flavor}$, given by $g_{IJ}$. These assumptions place constraints on possible choices of the symmetry groups.

We now define the YMS theory by the Lagrangian
\begin{equation}
\L_\text{YMS}=-\tfrac14g_{\a\b}F^\a_{\m\n}F^{\b\m\n}+\tfrac12g_{IJ}g_{ab}\Da_\m\p^{Ia}\Da^\m\p^{Jb}-\tfrac{1}{4} \lambda g_{IJ}g_{KL}g_{\a\b}f^\a_{ac}f^\b_{bd}\p^{Ia}\p^{Jb}\p^{Kc}\p^{Ld},
\label{YMS:lagrangian}
\end{equation}
where $\Da_\m\Phi^I\equiv\de_\m\Phi^I-\im[A_\m,\Phi^I]$, or equivalently $\Da_\m\p^{Ia}=\de_\m\p^{Ia}+f^a_{\a b}A^\a_\m\p^{Ib}$. The kinetic terms for the gauge and scalar fields are standard and fully symmetry-dictated. The only piece of the Lagrangian that requires some engineering is the quartic self-coupling of the scalars, with coupling strength $\lambda$: this was obtained by forming the commutators $[\Phi^I,\Phi^K]$ and $[\Phi^J,\Phi^L]$, and contracting them with the appropriate invariant flavor metrics.

The classical EoM for the scalar fields, following from the Lagrangian~\eqref{YMS:lagrangian}, reads
\begin{equation}
\boxed{\Da^2\p^{Ia}+\lambda g_{JK}g_{\a\b}g^{ab}f^\a_{bd}f^\b_{ce}\p^{Ic}\p^{Jd}\p^{Ke}=0,}
\label{YMS:EoMscalar}
\end{equation}
where $g^{ab}$ is, as usual, the matrix inverse of $g_{ab}$. (The same convention is used for the other metrics.) On the other hand, the EoM for the gauge field is
\begin{equation}
\boxed{\Da_\m F^{\a\m\n}+g_{IJ}g^{\a\b}g_{ac}f^c_{\b b}(\Da^\n\p^{Ia})\p^{Jb}=0.}
\label{YMS:EoMgauge}
\end{equation}
Equations~\eqref{YMS:EoMscalar} and~\eqref{YMS:EoMgauge} form the template for EoMs for all the theories that we consider, and will be used as a reference in the following sections. 


\subsection{Scattering amplitudes}
\label{subsec:YMS:amplitudesfromEoM}

It has been known for a long time that the solution of the classical EoM of a field theory in the presence of a background source acts as a generator of tree-level Green's functions~\cite{Boulware1968}. In ref.~\cite{Monteiro2011}, this approach was used to study the duality between scattering amplitudes in Yang-Mills theory and gravity, and in ref.~\cite{Cheung2021} it was further extended, among others, to certain EFTs of NG bosons. Our presentation below follows closely the latter.

Anticipating the duality with the NLSM, elaborated in detail in section~\ref{sec:flavorkinematics}, we will now further restrict the algebraic structure of the YMS theory by assuming that the metrics $g_{\a\b}$ and $g_{ab}$ and their inverses can be used to raise and lower indices on the ``structure constants'' $f^b_{\a a}$ and $f^\a_{ab}$ so that symbols with three subscripts are fully antisymmetric under permutations of the indices. Thus, for instance,
\begin{equation}
f^\a_{ab}=g^{\a\b}f_{\b ab}=g^{\a\b}f_{b\b a}=g^{\a\b}g_{bc}f^c_{\b a}.
\end{equation}
In addition, we will assume standard normalization of the kinetic terms for the scalar and gauge fields, which amounts to a choice of basis of fields and the corresponding matrices $Q_\a,Q_a$ so that $g_{\a\b}=\d_{\a\b}$, $g_{ab}=\d_{ab}$, and $g_{IJ}=\d_{IJ}$. 

We will be concerned solely with the scattering in the scalar sector of the YMS theory. In order to be able to extract the scattering amplitudes of the scalar particles corresponding to the fields $\p^{Ia}$, we deform the scalar EoM~\eqref{YMS:EoMscalar} by adding to its right-hand side a source, $J^{Ia}$. For the EoM~\eqref{YMS:EoMgauge} to represent a well-posed initial value problem for the gauge potential $A^\a_\m$, we impose the Lorenz gauge, $\de^\m A^\a_\m=0$. Upon expanding all the covariant derivatives, eqs.~\eqref{YMS:EoMscalar} and~\eqref{YMS:EoMgauge} can then be expressed jointly as
\begin{equation}
\begin{split}
\Box\p^{Ia}+2f^a_{\a b}A^{\a\m}\de_\m\p^{Ib}+f^a_{\a c}f^c_{\b b}A^\a_\m A^{\b\m}\p^{Ib}-\lambda\d_{JK}f^a_{\a c}f^\a_{bd}\p^{Ib}\p^{Jc}\p^{Kd}&=J^{Ia},\\
\Box A^\a_\m+2f^\a_{\b\g}A^{\b\n}\de_\n A^\g_\m-f^\a_{\b\g}A^\b_\n\de_\m A^{\g\n}+f^\a_{\b\eps}f^\eps_{\g\d}A^\b_\n A^{\g\n}A^\d_\m&\phantom{=}\\
-\d_{IJ}f^\a_{ab}(\de_\m\p^{Ia})\p^{Jb}+\d_{IJ}f^\a_{ac}f^c_{\b b}A^\b_\m\p^{Ia}\p^{Jb}&=0.
\end{split}
\label{YMS:EoMsource}
\end{equation}
These can be solved iteratively by formally expanding the fields,
\begin{equation}
\p^{Ia}=\sum_{n=1}^\infty\p^{(n)Ia},\qquad
A^\a_\m=\sum_{n=1}^\infty A^{(n)\a}_\m.
\label{YMS:iteration}
\end{equation}
The initial condition for the iteration is chosen in accord with the fact that we are only interested in scattering in the scalar sector of the YMS theory. Thus, we treat the sources $J^{Ia}$ as quantities of order one in the expansion, assuming that they create free on-shell scalar particles at spatial infinity. Since there are no asymptotic one-particle states corresponding to $A^\a_\m$, we set $A^{(1)\a}_\m=0$. The first two terms of the expansion of the scalar and gauge fields are then given by
\begin{equation}
\begin{aligned}
\p^{(1)Ia}&=\Box^{-1}J^{Ia},\qquad
& A^{(1)\a}_\m&=0,\\
\p^{(2)Ia}&=0,\qquad
& A^{(2)\a}_\m&=\d_{IJ}f^\a_{ab}\Box^{-1}[(\de_\m\p^{(1)Ia})\p^{(1)Jb}].
\end{aligned}
\end{equation}
Higher-order terms can in principle be obtained by direct iteration. It is however easier to interpret the individual contributions to the order-by-order solution of eq.~\eqref{YMS:EoMsource} diagrammatically.

To that end, we first recall that mathematically, the iterative solution for $\p^{Ia}$ represents the tree-level one-point function $\langle\p^{Ia}\rangle_J$ in presence of the external source. Its $\p^{(n)Ia}$ part accordingly collects all contributions of order $n$ in the sources. The individual contributions to the one-point function can be represented by rooted tree graphs, where one external leg (``root'') is singled out to correspond to the field $\p^{Ia}$ whose expectation value we are computing. All the remaining external legs (``leaves'') carry one factor of the external source each. We use the convention whereby the root is the leftmost external leg of the diagram, and a set of interaction vertices encoded in eq.~\eqref{YMS:EoMsource} defines possible branching of the diagram to the right. Each internal line of the diagram carries the propagator $\Box^{-1}$, which translates to $-1/p^2$ in momentum space. Equation~\eqref{YMS:EoMsource} encodes altogether seven different types of interaction vertices. These have the same topology as the interaction vertices of the ordinary perturbation theory. The difference is that in the rooted graph representation, the interaction vertices as well as the entire one-point function $\langle\p^{Ia}\rangle_J$ are only invariant under permutations of the leaf legs. The few specific vertices needed for the sample calculation below are listed in figure~\ref{fig:YMSvertices}.

\begin{figure}
\begin{align*}
\quad\smash{\feynSSSS{Ia}{Jb}{Kc}{Ld}}\qquad=-\lambda\bigl[&f_{\a ac}f^\a_{bd}(\d_{IJ}\d_{KL}-\d_{IL}\d_{JK})+f_{\a ab}f^\a_{dc}(\d_{IL}\d_{JK}-\d_{IK}\d_{JL})\\
&+f_{\a ad}f^\a_{cb}(\d_{IK}\d_{JL}-\d_{IJ}\d_{KL})\bigr]\\[7ex]
\feynSSG{Ia}{Jb}{\a\m}{p}\quad=&-2\im \d_{IJ}f_{\a ab}p^\m\qquad\qquad
\feynGSS{\a\m}{Ia}{Jb}{p}{q}\quad=-\im\d_{IJ}f_{\a ab}(p-q)^\m
\end{align*}
\caption{\emph{Sample interaction vertices of the rooted graph representation of the YMS theory. For each vertex, the root is the external leg on the left, whereas the external legs on the right are the leaves. The momenta \rev{$p,q$} carried by the leaves are always oriented towards the root.}}
\label{fig:YMSvertices}
\end{figure}

The desired scattering amplitude for $n$ scalar particles is extracted by first computing the one-point function $\langle\p^{Ia}\rangle_J$ to order $n-1$ in the source, that is, the $\p^{(n-1)Ia}$ part of the iterative solution of the EoM. Then one carries out the standard LSZ reduction whereby the sources on the leaf legs as well as all the external propagators are stripped off, and the momenta on all the $n$ external legs are taken to the mass shell. In this limit, one recovers the physical on-shell amplitude that has full invariance under permutations of~the~$n$~particles.

For the sake of a simple illustration, we now show how the rooted graph representation based on the classical EoM recovers the four-point amplitude of scalars in the YMS theory. This requires the third-order contribution to the one-point function, $\p^{(3)Ia}$, which is given by the diagrams\\[1ex]
\begin{align}
\label{YMS:V4graphs}
\parbox{20mm}{\begin{fmfgraph*}(20,20)
\fmfleft{l}
\fmfrightn{r}{3}
\fmf{plain}{r2,v,l}
\fmffreeze
\fmf{plain}{r1,v}
\fmf{plain}{r3,v}
\fmfv{label=$Ia$}{l}
\fmfv{label=$Kc$}{r2}
\fmfv{label=$Jb$}{r3}
\fmfv{label=$Ld$}{r1}
\fmfv{d.sh=circle,d.si=10thick}{v}
\end{fmfgraph*}}\qquad={}&\qquad\feynSSSS{Ia}{Jb}{Kc}{Ld}\\[5ex]
\notag
&+\qquad\parbox{30mm}{\begin{fmfgraph*}(30,20)
\fmfleft{l}
\fmfrightn{r}{3}
\fmf{plain}{r3,vu}
\fmf{phantom}{r1,vu}
\fmf{plain,tension=2}{vu,l}
\fmffreeze
\fmf{plain}{r2,vd}
\fmf{plain}{r1,vd}
\fmf{photon}{vu,vd}
\fmfv{label=$Ia$}{l}
\fmfv{label=$Kc$}{r2}
\fmfv{label=$Jb$}{r3}
\fmfv{label=$Ld$}{r1}
\fmfdot{vu,vd}
\end{fmfgraph*}}\qquad+\text{cycl.~perm.~of $Jb,Kc,Ld$}.\\[2ex]
\notag
\end{align}
Upon cutting off the external legs, this translates with the help of the Feynman rules listed in figure~\ref{fig:YMSvertices} to the amputated off-shell four-point vertex function of the YMS theory,
\begin{align}
\notag
V_4^\text{off-shell}={}&f_{\a ab}f^\a_{cd}\biggl[\lambda (\d_{IL}\d_{JK}-\d_{IK}\d_{LJ})+2\d_{IJ}\d_{KL}\frac{p_J\cdot(p_K-p_L)}{s_{IJ}}\biggr]\\
\label{YMS:V4resultoffshell}
&+f_{\a ac}f^\a_{db}\biggl[\lambda(\d_{IJ}\d_{KL}-\d_{IL}\d_{JK})+2\d_{IK}\d_{LJ}\frac{p_K\cdot(p_L-p_J)}{s_{IK}}\biggr]\\
\notag
&+f_{\a ad}f^\a_{bc}\biggl[\lambda(\d_{IK}\d_{LJ}-\d_{IJ}\d_{KL})+2\d_{IL}\d_{JK}\frac{p_L\cdot(p_J-p_K)}{s_{IL}}\biggr],
\end{align}
where we grouped the contact and exchange contributions carrying the same color structure (the three structures are of course not independent but subject to the Jacobi identity), and used the flavor indices to distinguish the different momenta. In addition, $s_{IJ}\equiv(p_I+p_J)^2$ are the standard Mandelstam variables; we use the convention that all the momenta $p_I,p_J,p_K,p_L$ are incoming. In the on-shell limit, our result can be rewritten in a way that makes the full permutation invariance manifest,
\begin{equation}
\begin{split}
V_4^\text{on-shell}={}&f_{\a ab}f^\a_{cd}\biggl[\lambda(\d_{IL}\d_{JK}-\d_{IK}\d_{LJ})-\d_{IJ}\d_{KL}\frac{(p_I-p_J)\cdot(p_K-p_L)}{s_{IJ}}\biggr]\\
&+f_{\a ac}f^\a_{db}\biggl[\lambda(\d_{IJ}\d_{KL}-\d_{IL}\d_{JK})-\d_{IK}\d_{LJ}\frac{(p_I-p_K)\cdot(p_L-p_J)}{s_{IK}}\biggr]\\
&+f_{\a ad}f^\a_{bc}\biggl[\lambda(\d_{IK}\d_{LJ}-\d_{IJ}\d_{KL})-\d_{IL}\d_{JK}\frac{(p_I-p_L)\cdot(p_J-p_K)}{s_{IL}}\biggr].
\end{split}
\label{YMS:V4resultonshell}
\end{equation}
This agrees, up to a conventional overall factor of $\im$, with the four-point on-shell amplitude of the YMS theory, derived using standard perturbation theory.


\section{Nonlinear sigma model}
\label{sec:NLSM}

\subsection{Algebraic structure}

Throughout the paper, we use the acronym NLSM as synonymous with a relativistic EFT for spontaneously broken internal symmetry at the leading order of the derivative expansion. In order to be able to write down its Lagrangian and EoM, we now have to review a necessary minimum of formalism. See chapters 7 and 8 of ref.~\cite{Brauner2024} for full detail. 

We initially consider an arbitrary symmetry-breaking pattern $G\to H$, only subject to the assumption that the unbroken subgroup $H$ is compact. We will use the notation $Q_{A,B,\dotsc}$ for the generators of $G$. In order to distinguish explicitly the generators of $H$, we will use $Q_{\a,\b,\dotsc}$ in accord with the notation introduced in section~\ref{sec:YMS}. Finally, we will use $Q_{a,b,\dotsc}$ for a chosen basis of broken generators.

The basic ingredient of the construction of the NLSM is the definition of action of the symmetry group $G$ on the coset space $G/H$. This can be brought by a suitable choice of local coordinates $\pi^a$ on $G/H$ to the form
\begin{equation}
U(\pi)\xrightarrow{g}U(\pi'(\pi,g))=gU(\pi)h(\pi,g)^{-1},\qquad g\in G,
\label{actionofG}
\end{equation}
where $U(\pi)$ is a unique representative of a given (left) coset of $H$ in $G$, parameterized by the coordinates $\pi^a$. Also, $h(\pi,g)\in H$ is inserted to project $gU(\pi)$ to the appropriate representative of its own coset.

Next, we introduce the Maurer-Cartan (MC) form, which will later play a central role in establishing a duality between the NLSM and the YMS theory. This is a Lie-algebra-valued 1-form,
\begin{equation}
\mathbf{\Omega}(\pi)\equiv-\im U(\pi)^{-1}\dd U(\pi)\equiv \mathbf{A}(\pi)+ \boldsymbol{\mc}(\pi),
\label{MCform}
\end{equation}
where $\mathbf{A}\equiv A^\a Q_\a$ and $\boldsymbol{\mc}\equiv\mc^a Q_a$ are the projections of the MC form to the subspaces of unbroken and broken generators, respectively. The two parts of the MC form transform under the action~\eqref{actionofG} of $G$ respectively as a gauge connection and as a set of covariant matter fields,
\begin{equation}
\mathbf{A} \xrightarrow{g}h \mathbf{A} h^{-1}-\im h\dd h^{-1},\qquad
\boldsymbol{\mc} \xrightarrow{g}h \boldsymbol{\mc} h^{-1},
\label{MCformtransfo}
\end{equation}
where $h$ stands for $h(\pi,g)$. Note that the broken part of the MC form is in a sense the simplest covariantly-transforming quantity built out of $\pi^a$, which themselves transform under $G$ in a complicated nonlinear fashion.  

The MC form satisfies the MC equation $\dd \mathrm{\Omega}^A=\tfrac12 f^A_{BC}\mathrm{\Omega}^B\w\mathrm{\Omega}^C$, where $f^A_{BC}$ are the structure constants of the symmetry group $G$. In terms of the unbroken and broken parts of the MC form, this splits as
\begin{equation}
\begin{split}
\dd A^\a& =\tfrac12f^\a_{\b\g}A^\b\w A^\g+\tfrac12 f^\a_{ab}\mc^a\w\mc^b,\\
\dd\mc^a&=f^a_{\a b} A^\a\w\mc^b+\tfrac12 f^a_{bc}\mc^b\w\mc^c.
\end{split}
\label{MCequations}
\end{equation}
The MC form encodes the data needed to characterize the geometry of the coset space. Its broken part, $\boldsymbol{\mc}$, constitutes a globally well-defined coframe on $G/H$. The unbroken part, $\mathbf{A}$, defines an $H$-connection. The first line of eq.~\eqref{MCequations} carries information about the curvature of this connection, whereas the second line thereof determines its torsion.

The NLSM as a Lagrangian field theory is constructed by replacing the local coordinates $\pi^a$ on $G/H$ with $G/H$-valued scalar fields, $\pi^a(x)$. This pulls the MC form back to a Lie-algebra-valued 1-form on the Minkowski spacetime, or equivalently a set of covariant vector fields $\mathrm{\O}^A_\m$, defined by $\mathrm{\O}^A\equiv \mathrm{\O}^A_\m\dd x^\m$. With the $H$-connection $A^\a_\m$ at hand, it is natural to define two covariantly transforming quantities, namely the field-strength of $A^\a_\m$ itself and the covariant derivative of $\mc^a_\m$,
\begin{equation}
F^\a_{\m\n}\equiv\de_\m A^\a_\n-\de_\n A^\a_\m-f^\a_{\b\g} A^\b_\m A^\g_\n,\qquad
\Da_\m\mc^a_\n\equiv\de_\m\mc^a_\n-f^a_{\a b} A^\a_\m\mc^b_\n.
\label{NLSM:GDmcdef}
\end{equation}
These are subject to the identities
\begin{equation}
F^\a_{\m\n}=f^\a_{ab}\mc^a_\m\mc^b_\n,\qquad
\Da_\m\mc^a_\n-\Da_\n\mc^a_\m=f^a_{bc}\mc^b_\m\mc^c_\n,
\label{NLSM:geomids}
\end{equation}
that follow from the two MC equations in eq.~\eqref{MCequations}.


\subsection{Equations of motion}

Turning to the Lagrangian of the NLSM, this is a quadratic function of the broken part of the MC form,
\begin{equation}
\L_\text{NLSM}=\frac 12\k_{ab}\mc^a_\m\mc^{b\m},
\label{NLSM:Lagrangian}
\end{equation}
where $\k_{ab}$ is a constant, symmetric and positive-definite matrix, invariant under the adjoint action of $H$. The corresponding EoM for the NG fields $\pi^a$ reads
\begin{equation}
\k_{ab}\Da_\m\mc^{b\m}-\k_{cd}f^d_{ab}\mc^b_\m\mc^{c\m}=0.
\label{NLSM:EoMgeneral}
\end{equation}

Everything said so far applies to an arbitrary choice of $G/H$. However, from now on we shall focus on coset spaces that are symmetric. In terms of the algebraic structure of the NLSM, the assumption of symmetry boils down to the vanishing of $f^c_{ab}$, so that $[Q_a,Q_b]=\im f^\a_{ab}Q_\a$. This reduces the EoM~\eqref{NLSM:EoMgeneral} to a covariant conservation law, $\Da_\m\mc^{a\m}=0$, which is entirely independent of the choice of the matrix coupling $\k_{ab}$. Moreover, the second identity in eq.~\eqref{NLSM:geomids} simplifies to $\Da_\m\mc^a_\n=\Da_\n\mc^a_\m$, which says that the $H$-connection on the coset space provided by the MC form is torsion-free. It now takes a simple manipulation to deduce a second-order differential equation for $\mc^a_\m$,
\begin{equation}
\Da^2\mc^a_\m=\Da^\n \Da_\n\mc^a_\m=\Da^\n \Da_\m\mc^a_\n=[\Da^\n,\Da_\m]\mc^a_\n=f^a_{\a b}F^\a_{\m\n}\mc^{b\n},
\label{NLSM:EoMderivation}
\end{equation}
where in the last step, we expressed a commutator of covariant derivatives in terms of the field strength $F^\a_{\m\n}$. The latter can finally be eliminated by using the first identity in eq.~\eqref{NLSM:geomids}, which leads to a second-order EoM for the MC form, 
\begin{equation}
\boxed{\Da^2\mc^a_\m+f^a_{\a b}f^\a_{cd}\mc^b_\n\mc^{c\n}\mc^d_\m=0,}
\label{NLSM:EoMscalar}
\end{equation}
that bears a remarkable resemblance to the EoM~\eqref{YMS:EoMscalar} for the scalar fields in the YMS theory.

To make the analogy with the YMS theory complete, we need a counterpart of the EoM~\eqref{YMS:EoMgauge} for the gauge field therein. The main difference between the YMS theory and the NLSM is that in the latter, the ``gauge field'' $A^\a_\m$ is composite and accordingly does not have independent dynamics. Indeed, the EoM~\eqref{NLSM:EoMgeneral}, or its descendant~\eqref{NLSM:EoMscalar}, establishes a well-defined initial value problem for the NG fields $\pi^a(x)$. This stems from the fact that we fixed the choice of the representative elements of the cosets, which by eq.~\eqref{MCformtransfo} amounts to fixing a gauge for $\mathbf{A}(\pi)$. There is an alternative approach that maintains the gauge freedom in the NLSM, known as the ``hidden local symmetry''~\cite{Bando1988a}. In this approach, one needs to augment eq.~\eqref{NLSM:EoMscalar} with an EoM for the gauge fields $A^\a_\m$, which constitute an independent set of degrees of freedom except for the covariant constraints~\eqref{NLSM:geomids}. This EoM can be deduced by taking the covariant derivative of the first identity in eq.~\eqref{NLSM:geomids} and using the covariant conservation law for $\mc^a_\m$ together with the symmetry of $\Da_\m\mc^a_\n$. We thus find
\begin{equation}
\boxed{\Da_\m F^{\a\m\n}+f^\a_{ab}(\Da^\n\mc^a_\m)\mc^{b\m}=0,}
\label{NLSM:EoMgauge}
\end{equation}
which is fully analogous to the EoM~\eqref{YMS:EoMgauge} for the YMS gauge fields.


\subsection{Scattering amplitudes}
\label{subsec:NLSM:amplitudesfromEoM}

Equations~\eqref{NLSM:EoMscalar} and~\eqref{NLSM:EoMgauge} together make a suitable starting point for the construction of tree-level scattering amplitudes in the NLSM, and for understanding their relationship to the scattering amplitudes in the scalar sector of the YMS theory. We will now review the amplitude construction treating NLSM as a standalone field theory, and return to the duality with the YMS theory in section~\ref{sec:flavorkinematics}.

Following the workflow as laid out in section~\ref{subsec:YMS:amplitudesfromEoM}, we next adopt the Lorenz gauge for the composite gauge field, $\de^\m A^\a_\m=0$, and expand the covariant derivatives in eqs.~\eqref{NLSM:EoMscalar} and~\eqref{NLSM:EoMgauge}. We also introduce a source $J^a$ for asymptotic one-particle states created by the NG fields $\pi^a$. Given that the MC form is linear in derivatives of the NG fields, we do so by adding to the right-hand side of eq.~\eqref{NLSM:EoMscalar} the term $\de_\m J^a$. Altogether, the second-order equations of motion of the NLSM thus become
\begin{equation}
\begin{split}
\Box\mc^a_\m-2f^a_{\a b}A^{\a\n}\de_\n\mc^b_\m+f^a_{\a c}f^c_{\b b}A^{\a}_{\n}A^{\b\n}\mc^b_\m+f^a_{\a b}f^\a_{cd}\mc^b_\n\mc^{c\n}\mc^d_\m&=\de_\m J^a,\\
\Box A^\a_\m-2f^\a_{\b\g}A^{\b\n}\de_\n A^\g_\m+f^\a_{\b\g} A^\b_\n\de_\m A^{\g\n}+f^\a_{\b\eps}f^\eps_{\g\d} A^\b_\n A^{\g\n} A^\d_\m&\\
-f^\a_{ab}(\de_\m\mc^a_\n)\mc^{b\n}+f^\a_{ac}f^c_{\b b}A^\b_\m\mc^a_\n\mc^{b\n}&=0.
\end{split}
\label{NLSM:EoMsource}
\end{equation}
These can again be solved iteratively. However, with the experience gathered in section~\ref{subsec:YMS:amplitudesfromEoM}, we proceed directly to the diagrammatic expansion of the one-point function $\langle\mc^a_\m\rangle_J$, which encodes tree-level scattering amplitudes of the NG modes. This is again organized in terms of rooted graphs. The root carries the MC form $\mc^a_\m$, whereas the leaves carry a derivative of the external source, $\de_\m J^a$, each. The branching of the trees is governed by the nonlinear terms in eq.~\eqref{NLSM:EoMsource}. A few sample interaction vertices, chosen to closely parallel those displayed in figure~\ref{fig:YMSvertices}, are shown in figure~\ref{fig:NLSMvertices}.

\begin{figure}
\begin{align*}
\quad\smash{\feynSSSS{a\m}{b\n}{c\k}{d\l}}\qquad=&-f^a_{\a b}f^\a_{cd}(\eta_{\m\l}\eta_{\n\k}-\eta_{\m\k}\eta_{\n\l})-f^a_{\a c}f^\a_{db}(\eta_{\m\n}\eta_{\k\l}-\eta_{\m\l}\eta_{\n\k})\\
&-f^a_{\a d}f^\a_{bc}(\eta_{\m\k}\eta_{\n\l}-\eta_{\m\n}\eta_{\k\l})\\[7ex]
\feynSSG{a\m}{b\n}{\a\l}{p}\quad=&-2\im \eta_{\m\n}f^a_{\a b}p_\l\qquad\qquad
\feynGSS{\a\m}{a\n}{b\l}{p}{q}\quad=\im \eta_{\n\l}f^\a_{ab}(p-q)_\m
\end{align*}
\caption{\emph{Sample interaction vertices of the rooted graph representation of the NLSM. For each vertex, the root is the external leg on the left, whereas the externals legs on the right are the leaves. The momenta \rev{$p,q$} carried by the leaves are always oriented towards the root.}}
\label{fig:NLSMvertices}
\end{figure}

To convert the one-point function $\langle\mc^a_\m\rangle_J$ to an on-shell scattering amplitude for the NG bosons requires a more careful inspection than in the YMS theory~\cite{Cheung2021}. The reason for this is that $\mc^a_\m$ itself is a composite operator, and moreover each leaf leg carries an extra derivative from $\de_\m J^a$. We start with the root leg. It is essential that the broken part of the MC form, $\mc^a_\m$, is a good interpolating field for the NG mode in the sense that its expansion in powers of the NG fields contains a linear term,\footnote{The form and normalization of the linear term is in principle arbitrary, but can be fixed by a suitable linear transformation of the NG fields.} $\mc^a_\m=\de_\m\pi^a+\mathcal O(\pi^2)$. By introducing an auxiliary momentum $q^\m$, we thus find that
\begin{equation}
\langle q^\m\mc^a_\m\rangle_J=\langle q\cdot \de\pi^a\rangle_J+\dotsb,
\end{equation}
where the ellipsis indicates contributions from the higher orders of the power expansion of $\mc^a_\m$, which drop out when the operator is coupled to an external single-particle state via the LSZ reduction. In momentum space, $q\cdot\de\pi^a$ becomes $\im p\cdot q\pi^a$, where $p^\m$ is the momentum carried by the particle on the root leg into the scattering domain. A similar but simpler argument applies to the leaf legs, where the extra derivative is likewise converted to $\im$ times the incoming momentum carried by the leg. The upshot of this discussion is that in order to convert the one-point function $\langle\mc^a_\m\rangle_J$ to a scattering amplitude, one has to cut off the propagators on the external legs, take the on-shell limit, and attach to the external root (R) and leaf (L) legs the polarization vectors
\begin{equation}
\varepsilon^\m_\text{root}(p)=-\frac{\im q^\m}{p\cdot q},\qquad
\varepsilon^\m_\text{leaf}(p)=\im p^\m.
\label{NLSM:polarizationvectors}
\end{equation}
Since the auxiliary momentum $q^\m$ can be chosen arbitrarily, all dependence on it in the final result for the on-shell amplitude must disappear. This is in practice a very useful check of consistency of the result for a scattering amplitude, and can be leveraged to reduce the computation effort and highlight specific properties of NLSM amplitudes, such as their vanishing and/or factorisation at certain kinematical configurations~\cite{Li:2024bwq}.

As a sample calculation, we now work out the amplitude for scattering of four NG bosons in the NLSM. This requires the part of the one-point function $\langle\mc^a_\m\rangle_J$ of third order in the source, represented diagrammatically by\\[1ex]
\begin{align}
\label{NLSM:V4graphs}
\parbox{20mm}{\begin{fmfgraph*}(20,20)
\fmfleft{l}
\fmfrightn{r}{3}
\fmf{plain}{r2,v,l}
\fmffreeze
\fmf{plain}{r1,v}
\fmf{plain}{r3,v}
\fmfv{label=$a\m$}{l}
\fmfv{label=$c\k$}{r2}
\fmfv{label=$b\n$}{r3}
\fmfv{label=$d\l$}{r1}
\fmfv{d.sh=circle,d.si=10thick}{v}
\end{fmfgraph*}}\qquad={}&\qquad\feynSSSS{a\m}{b\n}{c\k}{d\l}\\[5ex]
\notag
&+\qquad\parbox{30mm}{\begin{fmfgraph*}(30,20)
\fmfleft{l}
\fmfrightn{r}{3}
\fmf{plain}{r3,vu}
\fmf{phantom}{r1,vu}
\fmf{plain,tension=2}{vu,l}
\fmffreeze
\fmf{plain}{r2,vd}
\fmf{plain}{r1,vd}
\fmf{photon}{vu,vd}
\fmfv{label=$a\m$}{l}
\fmfv{label=$c\k$}{r2}
\fmfv{label=$b\n$}{r3}
\fmfv{label=$d\l$}{r1}
\fmfdot{vu,vd}
\end{fmfgraph*}}\qquad+\text{cycl.~perm.~of $b\n,c\k,d\l$}.\\[2ex]
\notag
\end{align}
Next, we cut off the external legs and append the polarization vectors~\eqref{NLSM:polarizationvectors}. This leads to a (still off-shell) four-point correlator $V_4^\text{off-shell}$, given by\footnote{\rev{Here and in similar expressions below, we lower an index on one of the structure constants via $f_{\a ab}=-f_{a\a b}=-\d_{ac}f^c_{\a b}$. This will make the full permutation invariance of the on-shell amplitude manifest, and also agrees with a similar convention adopted in the YMS theory.}}
\begin{equation}
\begin{split}
q\cdot p_aV_4^\text{off-shell}=&-f_{\a ab}f^\a_{cd}\biggl[(q\cdot p_d)(p_b\cdot p_c)-(q\cdot p_c)(p_d\cdot p_b)+2\frac{q\cdot p_b}{s_{ab}}(p_c\cdot p_d)p_b\cdot(p_c-p_d)\biggr]\\
&-f_{\a ac}f^\a_{db}\biggl[(q\cdot p_b)(p_c\cdot p_d)-(q\cdot p_d)(p_b\cdot p_c)+2\frac{q\cdot p_c}{s_{ac}}(p_d\cdot p_b)p_c\cdot(p_d-p_b)\biggr]\\
&-f_{\a ad}f^\a_{bc}\biggl[(q\cdot p_c)(p_d\cdot p_b)-(q\cdot p_b)(p_c\cdot p_d)+2\frac{q\cdot p_d}{s_{ad}}(p_b\cdot p_c)p_d\cdot(p_b-p_c)\biggr],
\end{split}
\label{NLSM:V4resultoffshell}
\end{equation}
where we used the NG flavor indices to label the momenta. This is already manifestly invariant under permutations of the leaf legs. 

Once we take the on-shell limit, two things happen simultaneously. First of all, the propagators get canceled by the momentum factors in the numerators. This is expected, since in a derivatively coupled EFT such as the NLSM, the on-shell four-point amplitude does not contain any physical factorization poles. Second, the dependence on the auxiliary momentum $q^\m$ should drop out. This is however not the case for the individual flavor structures corresponding to the three lines of eq.~\eqref{NLSM:V4resultoffshell}, as these are not mutually independent but related by the Jacobi identity,
\begin{equation}
f_{\a ab}f^\a_{cd}+f_{\a ac}f^\a_{db}+f_{\a ad}f^\a_{bc}=0.
\end{equation}
Writing the on-shell four-point amplitude generically as
\begin{equation}
V_4^\text{on-shell}=f_{\a ab}f^\a_{cd}V_4^{(b)}+f_{\a ac}f^\a_{db}V_4^{(c)}+f_{\a ad}f^\a_{bc}V_4^{(d)},
\end{equation}
the Jacobi identity makes the components $V_4^{(b)},V_4^{(c)},V_4^{(d)}$ ambiguous with respect to a shift by the same constant. We conclude that for a generic symmetric coset space $G/H$, there are only two independent ``partial'' amplitudes, corresponding to two arbitrarily chosen but linearly independent combinations $x_bV_4^{(b)}+x_cV_4^{(c)}+x_dV_4^{(d)}$ with $x_b+x_c+x_d=0$. We can for instance eliminate the $f_{\a ac}f^\a_{db}$ structure and express the on-shell limit of eq.~\eqref{NLSM:V4resultoffshell}~as
\begin{equation}
V_4^\text{on-shell}=f_{\a ab}f^\a_{cd}s_{ad}-f_{\a ad}f^\a_{bc}s_{ab}.
\end{equation}
By virtue of the Jacobi identity and the kinematical identities, satisfied by the Mandelstam variables, this is manifestly invariant under permutations of the four particles participating in the scattering process.


\subsection{Including gravity}
\label{subsec:NLSMg}

Before moving over to the next EFT with a higher soft degree, we would like to discuss one further generalization of the NLSM by coupling it to gravity (referred to as NLSM\textsubscript{g}). Following the minimal coupling procedure, this generalization is fully symmetry-dictated, and hence only introduces a single additional parameter, corresponding to the coupling strength of gravity. The coupling of the NLSM to gravity can be seen as similar to the coupling of scalars to gluons in the YMS theory. Both cases feature self-interactions between the propagating scalars with soft degrees $0$ and $1$, respectively, and the inclusion of spin-1 and spin-2 interactions is compatible with those soft degrees. As we will find in section~\ref{sec:flavorkinematics}, these gauge interactions are intrinsically necessary to realize the full mapping of the different theories, i.e.~the geometric duality that we are proposing.

Technically, the NLSM\textsubscript{g} is defined by coupling the NLSM~\eqref{NLSM:Lagrangian} minimally to a metric $g_{\m\n}$ and adding the Einstein-Hilbert action for the latter. Altogether, the action of the NLSM\textsubscript{g} thus reads
\begin{equation}
S_\text{NLSM\textsubscript{g}}=\int\dd^D\!x\,\sqrt{\abs{\det g}}\left(\frac{R}{16\pi G}+\frac12\k_{ab}g^{\m\n}\mc^a_\m\mc^b_\n\right),
\label{NLSMg:action}
\end{equation}
where $R$ is the Ricci scalar and $G$ the gravitational coupling. This gives a simple EoM for the metric, which separates the gravity and matter sectors,\footnote{This is a special case of a more general EoM, valid for Einsteinian gravity coupled to scalar matter whose Lagrangian density $\L_\text{mat}$ is linear in the inverse metric $g^{\m\n}$. That is the case for instance for any EFT of NG bosons at the leading, second order of the derivative expansion. In such cases, taking the trace of the Einstein equation leads readily to $R_{\m\n}=-16\pi G(\de\L_\text{mat}/\de g^{\m\n})$.}
\begin{equation}
R_{\m\n}=-8\pi G\k_{ab}\mc^a_\m\mc^b_\n,
\label{NLSMg:EoMgravity}
\end{equation}
where $R_{\m\n}$ is the Ricci tensor. We see that the symmetric matrix of coupling constants $\k_{ab}$ together with $G$ fixes the scale of coupling between the scalar sector and gravity. This will be important for matching the NLSM\textsubscript{g} to the DBI and sGal theories.

In order to be able to write down the EoM for the NG fields $\pi^a(x)$, we first introduce a doubly covariant derivative, which includes a minimal coupling to the gravity sector in the gauge-covariant derivative defined in eq.~\eqref{NLSM:GDmcdef},
\begin{equation}
\DD_\m\mc^a_\n\equiv\nabla_\m\mc^a_\n-f^a_{\a b}A^\a_\m\mc^b_\n=\de_\m\mc^a_\n-\Gamma^\l_{\m\n}\mc^a_\l-f^a_{\a b}A^\a_\m\mc^b_\n,
\label{NLSMgDD}
\end{equation}
where $\Gamma^\l_{\m\n}$ are the Christoffel symbols of the dynamical metric $g_{\m\n}$. The EoM then takes the form of a covariant conservation law,
\begin{equation}
\DD_\m\mc^{a\m}=0,
\end{equation}
which generalizes the EoM~\eqref{NLSM:EoMgeneral} of the NLSM (now restricted to symmetric coset spaces) by adding gravity. This can be converted to a second-order EoM by using the symmetry property of the covariant derivative, $\DD_\m\mc^a_\n=\DD_\n\mc^a_\m$, and following the same steps as in eq.~\eqref{NLSM:EoMderivation}, except that the commutator of covariant derivatives now receives a contribution from the Riemann curvature tensor,
\begin{equation}
[\DD_\m,\DD_\n]\mc^a_\l=-R^\k_{\phantom\k\l\m\n}\mc^a_\k-f^a_{\a b} F^\a_{\m\n}\mc^b_\l.
\end{equation}
Using finally eq.~\eqref{NLSMg:EoMgravity} to trade the Ricci tensor in the EoM for the MC form, we arrive at a generalization of eq.~\eqref{NLSM:EoMscalar} to the NLSM\textsubscript{g},
\begin{equation}
\DD^2\mc^a_\m+8\pi G\k_{bc}\mc^a_\n\mc^{b\n}\mc^c_\m+f^a_{\a b}f^\a_{cd}\mc^b_\n\mc^{c\n}\mc^d_\m=0.
\label{NLSMg:EoMscalar}
\end{equation}
It remains to write down a correspondingly generalized EoM for the unbroken part of the MC form, $A^\a_\m$. This follows as before by taking the covariant derivative of the first identity in eq.~\eqref{NLSM:geomids}, and generalizes the corresponding EoM~\eqref{NLSM:EoMgauge} in the NLSM by a trivial replacement of covariant derivatives,
\begin{equation}
\DD_\m F^{\a\m\n}+f^\a_{ab}(\DD^\n\mc^a_\m)\mc^{b\m}=0.
\label{NLSMg:EoMgauge}
\end{equation}


\section{Dirac-Born-Infeld theory}
\label{sec:DBI}

\subsection{Algebraic and geometric structure}

The DBI theory with $n$ flavors of scalars $\p^I$, $I=1,\dotsc,n$ is an EFT describing the fluctuations of a $D$-dimensional brane embedded in a flat $(D+n)$-dimensional spacetime. Below, we review the algebraic construction of the theory, following closely appendix A of ref.~\cite{Bogers2018b}, and give it a geometric interpretation. 

The starting point is the collection of all the symmetry generators: apart from the Poincar\'e generators $L_{\m\n}$ and $P_\m$, there are two sets of Lorentz-scalar generators, $Q_I$ and $Q_{IJ}$, and a collection of Lorentz vectors, $K_{\m I}$. Geometrically, the spontaneously broken generators $Q_I$ associated with the NG fields $\p^I$ can be viewed as generators of translations in the $n$ extra dimensions of the ambient spacetime. Similarly, the antisymmetric tensor $Q_{IJ}$ is associated with rotations in the extra dimensions, and finally $K_{\m I}$ generates rotations connecting the physical and the extra dimensions. 

The structure of the symmetry group is then fixed by the following commutators,
\begin{equation}
\begin{gathered}
[P_\m,K_{\n I}]=\im \eta_{\m\n}Q_I,\quad
[K_{\m I},K_{\n J}]=\im(g_{IJ}L_{\m\n}-\eta_{\m\n}Q_{IJ}),\quad
[K_{\m I},Q_J]=-\im g_{IJ}P_\m,\\
[Q_{IJ},K_{\m K}]=\im(g_{IK}K_{\m J}-g_{JK}K_{\m I}),\quad
[Q_{IJ},Q_K]=\im(g_{IK}Q_J-g_{JK}Q_I),\\
[Q_{IJ},Q_{KL}]=\im(g_{IK}Q_{JL}+g_{JL}Q_{IK}-g_{IL}Q_{JK}-g_{JK}Q_{IL}).
\end{gathered}
\label{DBI:commutators}
\end{equation}
All other commutators among the generators, not displayed above, are either fixed by Lorentz invariance or trivially zero. Taken together, these generators span the Lie algebra of the group $\gr{ISO}(D-1,n+1)$ or $\gr{ISO}(D+n-1,1)$, depending on the signature of the metric in the extra dimensions.\footnote{In order for the kinetic term of all the NG fields to have the correct sign and thus for the EFT to be perturbatively well-defined, the signature of the metric in all the extra dimensions must be the same. Note that the assumed existence of an ultraviolet completion rules out the version of the theory with extra timelike dimensions~\cite{Adams:2006sv}. Here we will treat both options on the same footing.} The convention used is such that $\eta_{\m\n}$ is the Minkowski metric in the physical spacetime with the ``mostly minus'' signature. In contrast, $g_{IJ}$ is the ambient metric in the extra dimensions with positive signature corresponding to spatial directions and negative signature to temporal directions. \rev{We can then set $g_{IJ}=\pm\d_{IJ}$ where the plus sign corresponds to the more physical case of extra spatial dimensions.}

In the ground state, the symmetry of the DBI theory is spontaneously broken down to $\gr{ISO}(D-1,1)\times\gr{O}(n)$, a combination of the physical Poincar\'e group and a group of flavor rotations, acting linearly on the NG fields. To set up the action of the symmetry group, we follow a generalization of the technique of nonlinear realizations, already utilized in section~\ref{sec:NLSM}, to spacetime symmetries~\cite{Volkov1973a,Ogievetsky1974a}. This is based on modding out all symmetries that are realized by linear transformations, which usually span a subset of the unbroken symmetries. The relevant coset space is thus respectively $\gr{ISO}(D-1,n+1)/[\gr{O}(D-1,1)\times\gr{O}(n)]$ or $\gr{ISO}(D+n-1,1)/[\gr{O}(D-1,1)\times\gr{O}(n)]$. To parameterize it, we need to introduce independent variables for all the nonlinearly realized symmetries: $x^\m$ for the physical spacetime translations $P_\m$, $\p^I$ for the extra-dimensional translations $Q_I$, and $\x^{\m I}$ for the rotations $K_{\m I}$. Together with the usual exponential parameterization, this gives the coset representative
\begin{equation}
U(x,\p,\x)\equiv e^{\im x^\m P_\m}e^{\im\p^IQ_I}e^{\im\x^{\m I}K_{\m I}}.
\label{DBI:coset}
\end{equation}
The group action is then defined as in eq.~\eqref{actionofG}, where $H$ is interpreted as the subgroup of linearly realized symmetries.

The basic geometric building blocks of the DBI theory can be obtained with the help of the MC form, defined in analogy with eq.~\eqref{MCform},
\begin{equation}
\mathbf{\O}\equiv-\im U^{-1}\dd U\equiv\tfrac12\mc^{\a\b}_L L_{\a\b}+\mc^\a_PP_\a+\mc^{\a I}_KK_{\a I}+\tfrac12\mc^{IJ}_QQ_{IJ}+\mc^I_QQ_I.
\end{equation}
Some of the components of the MC form are straightforward to evaluate explicitly in a closed form,
\begin{equation}
\mc^\a_P=\dd x^\b(\ch\Pi)_\b^{\phantom\b\a}-\dd\p^I\x_I^\b(\sh\Pi)_\b^{\phantom\b\a},\qquad
\mc^I_Q=\dd\p^I(\ch\Piinv)_J^{\phantom JI}-\dd x^\a\x_\a^J(\sh\Piinv)_J^{\phantom JI}.
\label{DBI:MC_PQ}
\end{equation}
Here $\sh$ and $\ch$ are shorthand notations for the functions $\sh(x)\equiv(\sinh\sqrt x)/\sqrt x$ and $\ch(x)\equiv\cosh\sqrt x$, and the matrices $\Pi_\m^{\phantom\m\n}$ and $\Piinv_I^{\phantom IJ}$, bilinear in the fields $\x^{\m I}$, are defined by
\begin{equation}
\Pi_\m^{\phantom\m\n}\equiv g_{IJ}\x^I_\m\x^{\n J},\qquad
\Piinv_I^{\phantom IJ}\equiv \eta_{\m\n}\x^\m_I\x^{\n J}.
\label{DBI:Pidef}
\end{equation}
The remaining pieces of the MC form can be found by a series expansion of the identity
\begin{equation}
\tfrac12\mc^{\a\b}_LL_{\a\b}+\mc^{\a I}_KK_{\a I}+\tfrac12\mc^{IJ}_QQ_{IJ}=-\im e^{-\im\x^{\m I}K_{\m I}}\dd e^{\im\x^{\m I}K_{\m I}}.
\label{DBI:MCformK}
\end{equation}

The only physical degrees of freedom of the DBI theory are the NG fields $\p^I$. The variables $x^\m,\x^{\m I}$ need to be included in the coset element~\eqref{DBI:coset} for appropriate nonlinear realization of the action of the generators $P_\m,K_{\m I}$. While $x^\m$ obviously play the role of spacetime coordinates, the fields $\x^{\m I}$ are auxiliary and can eventually be eliminated by a suitable inverse Higgs constraint (IHC)~\cite{Ivanov1975a}. Indeed, setting $\mc^I_Q=0$ gives a set of relations that can be solved for $\x^{\m I}$ in terms of derivatives of $\p^I$,
\begin{equation}
\de_\m\p^I=\x^J_\m\biggl(\frac{\sh\Piinv}{\ch\Piinv}\biggr)_J^{\phantom JI}=\biggl(\frac{\sh\Pi}{\ch\Pi}\biggr)_\m^{\phantom\m\n}\x^I_\n.
\label{DBI:IHC}
\end{equation}

We are now in the position to give a geometric interpretation to the individual components of the MC form (except for $\mc^I_Q$ that has been set to zero). First, $\mc^\a_P$ establishes a coframe $e^\a_\m$ on the fluctuating brane via $\mc^\a_P\equiv e^\a_\m\dd x^\m$. Using the IHC~\eqref{DBI:IHC}, the coframe can be cast as $e^\a_\m=(1/\ch\Pi)_\m^{\phantom\m\a}$. This gives in turn a metric $G_{\m\n}$ on the brane, induced by the ambient flat spacetime metric. Together with the corresponding Christoffel symbols of the first kind, the metric can be expressed solely in terms of the NG fields,
\begin{equation}
G_{\m\n}=\eta_{\a\b}e^\a_\m e^\b_\n=\eta_{\m\n}-g_{IJ}\de_\m\p^I\de_\n\p^J,\qquad
\Gamma_{\l\m\n}=-g_{IJ}\de_\l\p^I\de_\m\de_\n\p^J.
\label{DBI:metric}
\end{equation}
Second, the component $\mc^{\a\b}_L$ of the MC form equals \emph{minus} the conventional spin connection, associated with the Levi-Civita connection of the metric $G_{\m\n}$. Similarly, $\mc^{IJ}_Q$ defines a connection in the extra dimensions that allows one to build covariant derivatives with respect to the action of the unbroken $\gr{O}(n)$ symmetry, generated by $Q_{IJ}$. Finally, and most interestingly, $\mc^{\a I}_K\equiv\mc^{\a I}_{K\m}\dd x^\m$ can be contracted with the coframe, giving a rank-2 spacetime~tensor,
\begin{equation}
\K^I_{\m\n}\equiv \eta_{\a\b}e^\a_\m\mc^{\b I}_{K\n}.
\label{DBI:Kmunu}
\end{equation}
This has the interpretation as the extrinsic curvature of the brane. Its role is very similar to that of the broken MC components for the NLSM: it is the lowest-order (in a derivative expansion) quantity that transforms covariantly under the broken symmetry transformations. We will demonstrate that the EoM of the DBI theory takes the universal structure displayed in table~\ref{tab:bigpicture} when formulated in terms of this quantity.

To verify our interpretation of $\mc^{\a I}_K$, we work out the MC equations for the individual components of the MC form. These are easily deduced directly from the commutation relations among the various generators,
\begin{equation}
\begin{split}
\dd\mc^\a_P&=\eta_{\b\g}\mc^{\a\b}_L\w\mc^\g_P-g_{IJ}\mc^{\a I}_K\w\mc_Q^J,\\
\dd\mc_Q^I&=-g_{JK}\mc^{IJ}_Q\w\mc^K_Q+\eta_{\a\b}\mc^\a_P\w\mc^{\b I}_K,\\
\dd\mc^{\a I}_K&=\eta_{\b\g}\mc^{\a\b}_L\w\mc^{\g I}_K-g_{JK}\mc^{IJ}_Q\w\mc^{\a K}_K,\\
\dd\mc^{\a\b}_L&=-\eta_{\g\d}\mc^{\a\g}_L\w\mc^{\b\d}_L+g_{IJ}\mc^{\a I}_K\w\mc^{\b J}_K,\\
\dd\mc^{IJ}_Q&=g_{KL}\mc^{IK}_Q\w\mc^{JL}_Q-\eta_{\a\b}\mc^{\a I}_K\w\mc^{\b J}_K.
\end{split}
\label{DBI:MCequations}
\end{equation}
The MC equations encode much of the geometric structure of the DBI theory. The first line of eq.~\eqref{DBI:MCequations} states that, upon imposing the IHC $\mc^I_Q=0$, the spin connection $-\mc^{\a\b}_L$ has vanishing torsion with respect to the coframe $e^\a_\m$. The second line of eq.~\eqref{DBI:MCequations} asserts that, likewise upon using~$\mc^I_Q=0$, the tensor $\K^I_{\m\n}$~\eqref{DBI:Kmunu} is symmetric in its Lorentz indices. The third line of eq.~\eqref{DBI:MCequations} ensures that the exterior derivative of $\mc^{\a I}_K$, covariantized with respect to the action of both $L_{\m\n}$ and $Q_{IJ}$, vanishes. This implies the Codazzi equation,
\begin{equation}
\DD_\m\K^I_{\l\n}=\DD_\n\K^I_{\l\m},
\label{DBI:Codazzi}
\end{equation}
where we introduced a covariant derivative that receives contributions from both the Levi-Civita connection of the metric $G_{\m\n}$ (represented by the Christoffel symbols of the second kind, $\Gamma^\l_{\m\n}$) and the connection $\mc^{IJ}_Q$,
\begin{equation}
\DD_\m\K^I_{\n\l}\equiv\nabla_\m\K^I_{\n\l}+g_{JK}\mc^{IJ}_{Q\m}\K^K_{\n\l}=\de_\m\K^I_{\n\l}-\Gamma^\k_{\m\n}\K^I_{\k\l}-\Gamma^\k_{\m\l}\K^I_{\n\k}+g_{JK}\mc^{IJ}_{Q\m}\K^K_{\n\l}.
\label{DBI:covder}
\end{equation}
Together with the symmetry of $\K^I_{\m\n}$, the Codazzi equation~\eqref{DBI:Codazzi} ensures that the rank-3 spacetime tensor $\DD_\m\K^I_{\n\l}$ is fully symmetric in its Lorentz indices. Finally, the last two lines of eq.~\eqref{DBI:MCequations} connect the curvatures of the connections $\mc^{\a\b}_L$ and $\mc^{IJ}_Q$ to the extrinsic curvature $\K^I_{\m\n}$ of the brane. Namely, upon reformulation of the curvature 2-forms in terms of the Riemann curvature tensor of $G_{\m\n}$ and the corresponding field-strength tensor $F^{IJ}_{\phantom{IJ}\m\n}$ of $\mc^{IJ}_Q$, we arrive at the Gauss equations
\begin{equation}
R_{\k\l\m\n}=-g_{IJ}(\K^I_{\k\m}\K^J_{\l\n}-\K^I_{\k\n}\K^J_{\l\m}),\qquad
F^{IJ}_{\phantom{IJ}\m\n}=-G^{\k\l}(\K^I_{\k\m}\K^J_{\l\n}-\K^I_{\k\n}\K^J_{\l\m}).
\label{DBI:Gauss}
\end{equation}


\subsection{Equations of motion}
\label{subsec:DBIEoM}

Having given a thorough review of the algebraic and geometric aspects of the DBI theory, we now move on to its dynamics. The action of the DBI theory at the leading order of the derivative expansion is given simply by integration of the induced volume form on the brane, up to an overall scale that merely sets the units,
\begin{equation}
S_\text{DBI}=\int\dd^D\!x\,\sqrt{\abs{\det G}}.
\label{DBI:action}
\end{equation}
Using the identity $\d\abs{\det G}=\abs{\det G}G^{\m\n}\d G_{\m\n}$ to evaluate the variation of the action, we readily obtain the EoM in the form
\begin{equation}
\de_\m(\sqrt{\abs{\det G}}G^{\m\n}\de_\n\p^I)=0.
\end{equation}
With the explicit expression for the Christoffel symbols in eq.~\eqref{DBI:metric}, one finds that this is equivalent to $G^{\m\n}\de_\m\de_\n\p^I=0$. This form of the EoM is simpler, yet it is not manifestly covariant under the symmetries of the DBI theory, which obscures its geometric interpretation.

To make further progress, we digress and focus on the component $\mc^{\a I}_K$ of the MC form and the associated spacetime tensor $\K^I_{\m\n}$. It follows from eq.~\eqref{DBI:MCformK} that upon using the IHC~\eqref{DBI:IHC}, this tensor must take the schematic form $\K^I_{\m\n}=\de_\l\de_\n\p^JX^{\l I}_{\m J}(\de\p)$, where $X^{\l I}_{\m J}$ is some function of the first derivatives of $\p^I$. The symmetry of the extrinsic curvature, as required by the MC equations~\eqref{DBI:MCequations}, constrains the tensor structure of this function to $X^{\l I}_{\m J}(\de\p)=g^\l_\m\bar X^I_J(\de\p)$ so that $\K^I_{\m\n}=\de_\m\de_\n\p^J\bar X^I_J(\de\p)$. In addition, it is easy to compute the leading contribution to $\bar X^I_J$ in a series expansion in the NG fields, $\bar X^I_J(\de\p)=\d^I_J+\mathcal O((\de\p)^2)$. This ensures that $\bar X^I_J$ is an invertible matrix. As a consequence, the EoM of the DBI theory, which we previously found to take the form $G^{\m\n}\de_\m\de_\n\p^I=0$, is equivalent to
\begin{equation}
G^{\m\n}\K^I_{\m\n}=0.
\label{DBI:EoMtrace}
\end{equation}
This form of the EoM is manifestly covariant under all the symmetries of the DBI theory, and is therefore analogous to the covariant conservation law $\D_\m\mc^{a\m}=0$ in the NLSM. It is easy to interpret: the stationary points of the brane action~\eqref{DBI:action} are minimal surfaces in the ambient spacetime.

Using the EoM~\eqref{DBI:EoMtrace} together with the Codazzi equation~\eqref{DBI:Codazzi}, we next deduce that
\begin{equation}
0=\DD_\l(G^{\m\n}\K^I_{\m\n})=G^{\m\n}\DD_\l\K^I_{\m\n}=G^{\m\n}\DD_\n\K^I_{\m\l}=\DD^\m\K^I_{\m\l}.
\label{DBI:EoMcovcons}
\end{equation}
The extrinsic curvature is covariantly constant. This fact can be used further following the same line of reasoning as in the NLSM, cf.~eq.~\eqref{NLSM:EoMderivation},
\begin{equation}
\DD^2\K^I_{\m\n}=\DD^\l\DD_\l\K^I_{\m\n}=\DD^\l\DD_\m\K^I_{\l\n}=[\DD^\l,\DD_\m]\K^I_{\l\n}.
\label{DBI:EoMderivation}
\end{equation}
It remains to express the commutator of covariant derivatives in terms of the curvature tensors $R_{\k\l\m\n}$ and $F^{IJ}_{\phantom{IJ}\m\n}$. With the help of the Gauss equations~\eqref{DBI:Gauss}, one then arrives at a second-order EoM for the DBI theory with a contact quartic self-coupling,
\begin{equation}
\boxed{\DD^2\K^I_{\m\n}+g_{JK}\K^I_{\k\l}(2\K^{J\phantom\m\k}_{\phantom J\m}\K^{K\phantom\n\l}_{\phantom{K}\n}-\K^J_{\m\n}\K^{K\k\l})-g_{JK}\K^J_{\k\l}(\K^{K\phantom\m\k}_{\phantom K\m}\K^{I\phantom\n\l}_{\phantom I\n}+\K^{K\phantom\n\k}_{\phantom K\n}\K^{I\phantom\m\l}_{\phantom I\m})=0.}
\label{DBI:EoMscalar}
\end{equation}
While complicated, the interactions in this EoM for the extrinsic curvature are structurally analogous to the corresponding EoM for the YMS theory~\eqref{YMS:EoMscalar} and the NLSM~\eqref{NLSM:EoMscalar}.

In the special case of a single-flavor DBI theory, where the $\gr{O}(n)$ flavor indices are absent, the EoM~\eqref{DBI:EoMscalar} simplifies to
\begin{equation}
\nabla^2\K_{\m\n}-g\K_{\k\l}\K^{\k\l}\K_{\m\n}=0\qquad\text{(single-flavor DBI theory)},
\end{equation}
where $g$ is the sole matrix element of the metric $g_{IJ}$ in the extra dimension. Also other ingredients of the DBI theory can be given a simple, more explicit form in the single-flavor case. Thus, the metric in eq.~\eqref{DBI:metric} can be easily inverted, leading in turn to the Christoffel symbols of the second kind,
\begin{equation}
G^{\m\n}=\eta^{\m\n}+\frac{g\de^\m\phi\de^\n\phi}{1-g(\de\phi)^2},\qquad
\Gamma^\l_{\m\n}=-\frac{g\de^\l\phi\de_\m\de_\n\phi}{1-g(\de\phi)^2}.
\end{equation}
The extrinsic curvature~\eqref{DBI:Kmunu} likewise reduces to
\begin{equation}
\K_{\m\n}=\frac{\de_\m\de_\n\phi}{\sqrt{1-g(\de\phi)^2}}.
\end{equation}
The primary EoM~\eqref{DBI:EoMtrace} of the single-flavor DBI theory is then equivalent to
\begin{equation}
\Box\phi+\frac{g\de^\m\phi\de^\n\phi\de_\m\de_\n\phi}{1-g(\de\phi)^2}=0,
\end{equation}
which is easily checked directly starting from the action~\eqref{DBI:action}. The DBI symmetry generated by $K_\m$ acts as a field-dependent diffeomorphism on the spacetime coordinates, shifting them by $\delta x^\m=gc^\m\phi\equiv\zeta^\m$, where $c^\m$ is an infinitesimal parameter. The transformation of the scalar $\phi$ includes an additional shift as appropriate for a NG field,
\begin{equation}
\delta\phi=c^\mu x_\mu-\zeta^\mu\de_\mu\phi.
\end{equation}
On the other hand, the extrinsic curvature $\K_{\m\n}$ is guaranteed by construction to transform linearly under the induced coordinate diffeomorphism,
 \begin{align}
  \delta\K_{\mu\nu} =  -\zeta^\l \partial_\l\K_{\mu\nu} -\partial_\mu \zeta^\l\K_{\l \nu} - \partial_\nu \zeta^\l \K_{\mu \l}.
 \end{align}

Returning to the multiflavor case, we would eventually like to encode the dynamics of the DBI theory in terms of a set of second-order EoMs that can be used to establish a diagrammatic representation of the scattering amplitudes of the theory. To that end, we need the corresponding equations for the gauge sector as well. As to the gauge connection of the linearly realized $\gr{O}(n)$ flavor symmetry, $\mc^{IJ}_Q$, we simply take the covariant divergence of the field-strength tensor 
$F^{IJ}_{\phantom{IJ}\m\n}$ in eq.~\eqref{DBI:Gauss}. This leads, upon using the covariant conservation law for $\K^I_{\m\n}$ and the Codazzi equation~\eqref{DBI:Codazzi}, to 
\begin{equation}
\boxed{\DD_\m F^{IJ\m\n}-(\DD^\n\K^I_{\m\l})\K^{J\m\l}+(\DD^\n\K^J_{\m\l})\K^{I\m\l}=0,}
\label{DBI:EoMgauge}
\end{equation}
which is the analog of eq.~\eqref{YMS:EoMgauge} in the YMS theory and eq.~\eqref{NLSM:EoMgauge} in the NLSM.

In addition, the symmetric tensor field $\K^I_{\m\n}$ is coupled to a composite gravitational sector defined by the metric $G_{\m\n}$. Here we follow in spirit the philosophy we used previously in section~\ref{sec:NLSM} for the NLSM. Therein, we gave the composite gauge field $A^\a_\m$ its own dynamics by maintaining the freedom to choose the representative $U(\pi)$ of cosets of $H$ in $G$, and thus treating the unbroken subgroup $H$ as a gauge group acting on $G$-valued matrix fields. In the context of the DBI theory, this translates to the freedom to choose the parameterization of the embedding of the brane in the ambient space. This gives the metric $G_{\m\n}$ the status of an independent dynamical variable. The corresponding Einstein equation constitutes a relation for the Ricci tensor, $R_{\m\n}$. By taking the trace of the Gauss equation~\eqref{DBI:Gauss} and using the EoM~\eqref{DBI:EoMtrace} which dictates that the trace of the extrinsic curvature of the brane vanishes, we arrive at
\begin{equation}
\boxed{R_{\m\n}=g_{IJ}G^{\k\l}\K^I_{\k\m}\K^J_{\l\n}.}
\label{DBI:EoMgravity}
\end{equation}


\subsection{Scattering amplitudes}
\label{subsec:DBI:amplitudesfromEoM}

Equations~\eqref{DBI:EoMscalar}, \eqref{DBI:EoMgauge} and~\eqref{DBI:EoMgravity} together constitute a closed set of second-order partial differential equations for the dynamical variables $\K^I_{\m\n}$, $\mc^{IJ}_{Q\m}$ and $G_{\m\n}$ (with the associated Levi-Civita connection). We will now illustrate the use of these equations for computation of scattering amplitudes by working out the four-point amplitude of the DBI theory.

This amplitude is obtained by expanding the one-point function $\langle\K^I_{\m\n}\rangle_J$ to the third order in external sources $J^I$ for the DBI scalars $\p^I$. It can be represented by the same diagrams as in eqs.~\eqref{YMS:V4graphs} and~\eqref{NLSM:V4graphs}, where the propagators in the exchange diagrams carry either the composite gauge field $\mc^{IJ}_{Q\m}$ or the perturbation $h_{\m\n}$ of the composite metric $G_{\m\n}$ around a flat background, defined by
\begin{equation}
G_{\m\n}\equiv \eta_{\m\n}+h_{\m\n}.
\end{equation}
This means that to the order we are interested in (relevant for four-point amplitudes), it is sufficient to linearize the EoMs in the two composite gauge fields. We use the Lorenz gauge for the flavor gauge field, $\de^\m\mc^{IJ}_{Q\m}=0$. As to the metric perturbation, we start by writing down the linearized Christoffel symbols of $G_{\m\n}$,\rev{
\begin{equation}
\Gamma^\l_{\m\n}\approx\frac12\eta^{\l\k}(\de_\m h_{\k\n}+\de_\n h_{\m\k}-\de_\k h_{\m\n}).
\end{equation}}
The corresponding linearized Ricci tensor is
\begin{equation}
R_{\m\n}\approx\de_\l\Gamma^\l_{\n\m}-\de_\n\Gamma^\l_{\l\m}\approx\frac12(\de_\m\de_\l h_\n^{\phantom\n\l}+\de_\n\de_\l h_\m^{\phantom \m\l}-\de_\m\de_\n h-\Box h_{\m\n}),
\end{equation}
where $h\equiv h^\m_{\phantom\m\m}$ and indices in the linearized equations are raised and lowered using the flat metric $\eta_{\m\n}$. At this stage, we impose the harmonic gauge, $G^{\m\n}\Gamma^\l_{\m\n}=0$, which implies that $\de_\n h_\m^{\phantom\m\n}\approx(1/2)\de_\m h$. In the harmonic gauge, the linearized Ricci tensor and the Christoffel symbols satisfy the useful identities
\begin{equation}
R_{\m\n}\approx-\frac12\Box h_{\m\n},\qquad
\de^\m\Gamma^\l_{\m\n}\approx\frac12\Box h^\l_{\phantom\l\n}.
\end{equation}

At the linear order in the gauge fields, eqs.~\eqref{DBI:EoMgauge} and~\eqref{DBI:EoMgravity} now take the simple form
\begin{equation}
\Box\mc^{IJ}_{Q\m}\approx(\de_\m\K^I_{\n\l})\K^{J\n\l}-(\de_\m\K^J_{\n\l})\K^{I\n\l},\qquad
\Box h_{\m\n}\approx-2g_{IJ}\K^I_{\m\l}\K^{J\phantom \n\l}_{\phantom J\n}.
\label{DBI:linearizedEoMgauge}
\end{equation}
The EoM~\eqref{DBI:EoMscalar} for the matter field $\K^I_{\m\n}$ takes some more effort to expand. Upon using the EoM for $h_{\m\n}$ to remove terms containing $\Box h_{\m\n}$, it can be simplified to
\begin{equation}
\begin{split}
\Box\K^I_{\m\n}\approx&-2g_{JK}\K^I_{\k\l}\K^{J\phantom\m\k}_{\phantom J\m}\K^{K\phantom\n\l}_{\phantom{K}\n}+g_{JK}\K^I_{\k\l}\K^J_{\m\n}\K^{K\k\l}-2g_{JK}\mc^{IJ}_{Q\l}\de^\l\K^K_{\m\n}\\
&+h^{\k\l}\de_\k\de_\l\K^I_{\m\n}+[(\de_\m h^\k_{\phantom\k\l}+\de_\l h^\k_{\phantom\k\m}-\de^\k h_{\m\l})\de^\l\K^I_{\k\n}+(\m\leftrightarrow\n)].
\end{split}
\end{equation}
The first line corresponds to the contact four-point contributions and the exchange of the composite flavor gauge bosons, whereas the second line amounts to the exchange of the composite graviton. Upon using eq.~\eqref{DBI:linearizedEoMgauge} to substitute for the gauge fields, we finally obtain an expression for $\K^{(3)I}_{\m\n}$ in terms of $\K^{(1)I}_{\m\n}$ using the notation introduced in eq.~\eqref{YMS:iteration},
\begin{align}
\notag
\Box\K^I_{\m\n}\approx&-2g_{JK}\K^I_{\k\l}\K^{J\phantom\m\k}_{\phantom J\m}\K^{K\phantom\n\l}_{\phantom{K}\n}+g_{JK}\K^I_{\k\l}\K^J_{\m\n}\K^{K\k\l}\\
\notag
&-2g_{JK}(\de^\l\K^K_{\m\n})\Box^{-1}\bigl[(\de_\l\K^I_{\r\s})\K^{J\r\s}-(\de_\l\K^J_{\r\s})\K^{I\r\s}\bigr]\\
\label{DBI:V4}
&-2g_{JK}(\de^\k\de^\l\K^I_{\m\n})\Box^{-1}(\K^J_{\k\s}\K^{K\phantom\l\s}_{\phantom K\l})\\
\notag
&-2g_{JK}\bigl\{(\de^\l\K^I_{\k\n})\Box^{-1}[\de_\m(\K^{J\k\s}\K^K_{\l\s})+\de_\l(\K^{J\k\s}\K^K_{\m\s})-\de^\k(\K^J_{\m\s}\K^{K\phantom\l\s}_{\phantom K\l})]+(\m\leftrightarrow\n)\bigr\}.
\end{align}

To convert this to an actual on-shell scattering amplitude, we use the fact that $\K^I_{\m\n}=\de_\m\de_\n\p^I+\mathcal O(\p^2)$. The one-point function of $\p^I$ can thus be extracted from the one-point function of $\K^I_{\m\n}$ by introducing an auxiliary momentum $q^\m$ and contracting $\langle K^I_{\m\n}\rangle_J$ with $q^\m q^\n$.\footnote{It is also possible to adopt a more general convention and introduce two auxiliary momenta $q^\m_1,q^\m_2$, one for each Lorentz index of $\K^I_{\m\n}$. However, our setup is simpler in that it will not require explicit symmetrization in the two indices.} The scattering amplitude is then obtained from $\langle K^I_{\m\n}\rangle_J$ by cutting off the external propagators, taking the on-shell limit, and decorating the root and leaf legs with the respective polarization tensors
\begin{equation}
\varepsilon^{\m\n}_\text{root}(p)=-\frac{q^\m q^\n}{(p\cdot q)^2},\qquad
\varepsilon^{\m\n}_\text{leaf}(p)=-p^\m p^\n.
\label{DBI:polarizationvectors}
\end{equation}
Upon some further manipulation, we find that the singularities coming from the gauge field and graviton propagators as well as all dependence on $q^\m$ drop out as they should. The final result for the four-point on-shell amplitude of the DBI theory is
\begin{equation}
\rev{V_4^\text{on-shell}=-\frac12(\d_{IJ}g_{KL}s_{IK}s_{IL}+\d_{IK}g_{LJ}s_{IL}s_{IJ}+\d_{IL}g_{JK}s_{IJ}s_{IK}).}
\label{DBI:amplitude}
\end{equation}
\rev{In spite of appearance, this expression is invariant under permutations of all four particles as it should.}

Our calculation is obviously not the most economic way to derive the four-point amplitude of the DBI theory; it is easy to check eq.~\eqref{DBI:amplitude} by directly expanding the action~\eqref{DBI:action} in terms of the NG fields $\p^I$ and using standard perturbation theory. However, the above exercise can be viewed as a proof of concept, demonstrating explicitly that the computation of the scattering amplitudes in the DBI theory based on the classical EoM is feasible.


\section{Special Galileon theory}
\label{sec:Galileon}

\subsection{Algebraic and geometric structure}

The special Galileon is a single-scalar theory that realizes the highest possible soft degree, and as such is a natural extension of the NLSM and the DBI theory. Its symmetry algebra was pinned down in ref.~\cite{Hinterbichler2015a}, soon after the prediction of the existence of the theory based on the scaling of its scattering amplitudes in the soft limit~\cite{Cheung2015a}. It includes, in addition to the Poincar\'e generators $L_{\m\n}$ and $P_\m$, a Lorentz vector $K_\m$, a traceless symmetric tensor $S_{\m\n}$, and a scalar generator $Q$. The commutators of the sGal algebra that are nonvanishing and are not fixed by Lorentz invariance can be summarized as
\begin{equation}
\begin{split}
[P_\m,K_\n]&=\im \eta_{\m\n}Q,\\
[S_{\m\n},S_{\k\l}]&=\im\s(\eta_{\m\l}L_{\n\k}+\eta_{\n\k}L_{\m\l}+\eta_{\m\k}L_{\n\l}+\eta_{\n\l}L_{\m\k}),\\
[S_{\m\n},P_\l]&=\im\biggl(\eta_{\m\l}K_\n+\eta_{\n\l}K_\m-\frac2D \eta_{\m\n}K_\l\biggr),\\
[S_{\m\n},K_\l]&=\im\s\biggl(\eta_{\m\l}P_\n+\eta_{\n\l}P_\m-\frac2D \eta_{\m\n}P_\l\biggr).
\end{split}
\label{sGal:commutators}
\end{equation}
Here $\s=\pm1$ distinguishes between two different versions of the sGal theory (similar to the signature of the internal metric $g_{IJ}$ for the DBI theory). In case of $\s=+1$, the symmetry of the sGal theory coincides with the affine group of the physical Minkowski spacetime~\cite{Roest2021}. In particular, the set of generators $\{L_{\m\n},S_{\m\n}\}$ spans the Lie algebra of $\gr{SL}(D,\mathbbm{R})$, the group of all real unimodular linear transformations of the Minkowski coordinates. On the other hand, for $\s=-1$, the generators $\{L_{\m\n},S_{\m\n}\}$ span the Lie algebra of $\gr{SU}(D-1,1)$, and the sGal theory naturally arises from the K\"ahler geometry of a $D$-dimensional complex space~\cite{Novotny2017a}. 

The algebraic approach based on the technique of nonlinear realizations was first applied to the sGal theory in ref.~\cite{Bogers2018b} in the special case of $D=4$ spacetime dimensions, and generalized to arbitrary $D$ in ref.~\cite{GarciaSaenz2019}. Here we will review the algebraic construction and match it explicitly to the geometric structure underlying the sGal theory. As in the case of the DBI theory in section~\ref{sec:DBI}, this will equip us automatically and naturally with all the building blocks needed to derive the classical EoM in a form that matches those of the YMS theory~\eqref{YMS:EoMscalar}, the NLSM~\eqref{NLSM:EoMscalar}, and the DBI theory~\eqref{DBI:EoMscalar}.

The scalar generator $Q$ is associated with the sole physical degree of freedom of the sGal theory: the NG field $\p$. However, only the Lorentz subgroup generated by $L_{\m\n}$ is linearly realized. In order to be able to work out the action of the whole sGal symmetry group, we therefore need to introduce additional variables that take account of the nonlinear realization of spacetime translations (that is the vector of spacetime coordinates $x^\m$) as well as of the transformations generated by $K_\m$ (an auxiliary vector field $\x^\m$) and by $S_{\m\n}$ (an auxiliary traceless symmetric tensor field $\c^{\m\n}$). Altogether, the coset space carrying a nonlinear realization of the symmetry group can be parameterized as
\begin{equation}
U(x,\p,\x,\c)\equiv e^{\im x^\m P_\m}e^{\im\p Q}e^{\im\x^\m K_\m}e^{(\im/2)\c^{\m\n}S_{\m\n}}.
\end{equation}
Following the same steps as for the DBI theory, we next introduce the MC form and its individual components,
\begin{equation}
\mathbf{\Omega}\equiv-\im U^{-1}\dd U\equiv\tfrac12\mc^{\a\b}_LL_{\a\b}+\mc^\a_PP_\a+\tfrac12\mc^{\a\b}_SS_{\a\b}+\mc^\a_KK_\a+\mc_QQ.
\end{equation}
In contrast to the DBI theory, the commutation relations~\eqref{sGal:commutators} are simple enough so that all the components of the MC form can be evaluated in a closed form,
\begin{align}
\notag
\mc^{\a\b}_L&=\dd\c^{\m\n}\left\{B^{-1}\left[\cosh(\sqrt\s B)-\un\right]\right\}^{\a\b}_{\m\n},
&\mc^\a_P&=\dd x_\b\cosh(\sqrt\s\c)^{\a\b}+\sqrt\s\dd \x_{\b}\sinh(\sqrt\s\c)^{\a\b},\\
\notag
\mc^{\a\b}_S&=\dd\c^{\m\n}\left[(\sqrt\s B)^{-1}\sinh(\sqrt\s B)\right]^{\a\b}_{\m\n},
&\mc^\a_K&=\frac1{\sqrt\s}\dd x_\b \sinh(\sqrt\s\c)^{\a\b}+\dd \x_{\b}\cosh(\sqrt\s\c)^{\a\b},\\
\mc_Q&=\dd\p-\dd x^\a\x_\a,&&
\label{sGal:MCform}
\end{align}
with the shorthand notation $B^{\a\b}_{\m\n}\equiv \c^\a_\m\eta^\b_\n-\c^\b_\n\eta^\a_\m$.

To eliminate the auxiliary degrees of freedom $\x^\m$ and $\c^{\m\n}$, we need a set of IHCs, covariant under all the symmetries of the sGal theory. The first part is easy: setting $\mc_Q=0$ eliminates $\x^\m$ in favor of $\p$,
\begin{equation}
\x_\m=\de_\m\p.
\label{sGal:IHCphi}
\end{equation}
The elimination of $\c^{\m\n}$ is more subtle, and we will use $\mc^\a_K$ for the purpose. To that end, we however first have to turn the latter into a rank-2 spacetime tensor and then take a traceless symmetric part. As a first step towards this goal, we use $\mc^\a_P$ to define a coframe $e^\a_\m$ via $\mc^\a_P\equiv e^\a_\m\dd x^\m$. Next, it is convenient to switch temporarily to a matrix notation. Thus, we interpret $e^\a_\m$ as the matrix elements $\mathbbm{E}^\a_{\phantom\a\m}$ of a matrix $\mathbbm{E}$, trade $\c^{\m\n}$ for the matrix elements $\c^\m_{\phantom\m\n}$ of a matrix-valued field, $\c$, and treat $\de_\m\x^\n=\de_\m\de^\n\p$ as the matrix elements $\Xi^\n_{\phantom\n\m}$ of a matrix $\Xi$. Finally, we introduce the matrix $\mathbbm{K}$ whose matrix elements are the components of $\mc^\a_K$, that is, $\mc^\a_K\equiv\mathbbm{K}^\a_{\phantom\a\m}\dd x^\m$. With this notation, the relations for $\mc^\a_P$ and $\mc^\a_K$ in eq.~\eqref{sGal:MCform} can be written as
\begin{equation}
\mathbbm{E}=\cosh(\sqrt\s\c)+\sqrt\s\sinh(\sqrt\s\c)\cdot\Xi,\qquad
\mathbbm{K}=\frac1{\sqrt\s}\sinh(\sqrt\s\c)+\cosh(\sqrt\s\c)\cdot\Xi.
\label{sGal:IHCaux}
\end{equation}

Here comes the key step~\cite{GarciaSaenz2019}. The matrices $\mathbbm{K}$ and $\mathbbm{E}$ together define a set of covariant derivatives $\Da_\b\x^\a$ of the auxiliary vector field via $\mathbbm K^\a_{\phantom\a\m}\equiv(\Da_\b\x^\a)\mathbbm E^\b_{\phantom\b\m}$. Treating the latter again as the elements of a matrix, $(\Da\x)^\a_{\phantom\a\b}$, we can easily invert the relation to $\Da\x=\mathbbm K\cdot\mathbbm E^{-1}$. Our next IHC, used to eliminate the tensor field $\chi^{\m\n}$, is that the traceless symmetric part of the rank-2 covariant tensor $\Da_\a\x_\b\equiv \eta_{\b\g}\Da_\a\x^\g$ vanishes. However, as we will show below, this tensor already is automatically symmetric for fields satisfying the IHC~\eqref{sGal:IHCphi}. Thus, our new IHC really says that $\Da_\a\x_\b\propto \eta_{\a\b}$, which is equivalent to
\begin{equation}
\Da\x=\X\un\quad\text{where}\quad\X\equiv\frac1D\tr(\Da\x)\qquad\Leftrightarrow\qquad
\mathbbm{K}^\a_{\phantom\a\m}=\X\mathbbm{E}^\a_{\phantom\a\m}.
\label{sGal:IHCxi}
\end{equation}
This determines $\c$ uniquely as a function of $\Xi$ and $\tr\Xi$, which guarantees that the matrices $\c$ and $\Xi$ commute with each other. With this hindsight, it is easy to solve explicitly for $\c$,
\begin{equation}
\sqrt\s\c=-\arctanh(\sqrt\s\Xi)+\frac1D\un\tr\arctanh(\sqrt\s\Xi).
\label{sGal:IHCchi}
\end{equation}
Together with eq.~\eqref{sGal:IHCphi}, this eliminates both sets of auxiliary degrees of freedom from the sGal theory. Using eq.~\eqref{sGal:IHCaux}, the coframe $e^\a_\m$ can now be expressed solely in terms of $\p$ in the matrix form
\begin{equation}
\mathbbm E=\sqrt{\frac{\un-\s\Xi^2}{1-\s\X^2}},\qquad
\X=\frac1{\sqrt\s}\tanh\left[\frac1D\tr\arctanh(\sqrt\s\Xi)\right].
\label{sGal:Xdef}
\end{equation}

To make further progress towards a geometric interpretation of the algebraic structure of the sGal theory, we utilize the MC equations,
\begin{equation}
\begin{split}
\dd\mc^{\a\b}_L&=-\eta_{\g\d}\mc^{\a\g}_L\w\mc^{\b\d}_L+\s \eta_{\g\d}\mc^{\a\g}_S\w\mc^{\b\d}_S,\\
\dd\mc^{\alpha\beta}_S&=\eta_{\g\d}(\mc^{\a\g}_L\w\mc^{\b\d}_S+\mc^{\b\g}_L\w\mc^{\a\d}_S),\\
\dd\mc^\a_P&=\eta_{\b\g}\mc^{\a\b}_L\w\mc^\g_P+\s \eta_{\b\g}\mc^{\a\b}_S\w\mc^{\g}_K,\\
\dd\mc^\a_K&=\eta_{\b\g}\mc^{\a\b}_L\w\mc^\g_K+\eta_{\b\g}\mc^{\a\b}_S\w\mc^{\g}_P,\\
\dd\mc_Q&=\eta_{\a\b}\mc^\a_P\w\mc^\b_K.
\end{split}
\label{sGal:MCequations}
\end{equation}
The last of these implies upon using the IHC $\mc_Q=0$ that the rank-2 spacetime tensor $\eta_{\a\b}e^\a_\m\mathbbm{K}^\b_{\phantom\b\n}$ is symmetric. This is equivalent to the symmetry of $\Da_\a\x_\b$, which verifies the assumption that played an important role in the derivation of eq.~\eqref{sGal:IHCxi}. With the help of the latter, the third and fourth line of eq.~\eqref{sGal:MCequations} can be rewritten equivalently as
\begin{equation}
\D\mc^\a_P=\s\X \eta_{\b\g}\mc^{\a\b}_S\w\mc^\g_P,\qquad
\D\X\w\mc^\a_P=(1-\s\X^2)\eta_{\b\g}\mc^{\a\b}_S\w\mc^\g_P,
\label{sGal:MCrem}
\end{equation}
where $\D$ is an exterior derivative, covariant with respect to the spin connection $-\mc^{\a\b}_L$. The fact that $\D\mc^\a_P$ is nonzero indicates that the spin connection has a nonvanishing torsion with respect to the coframe $e^\a_\m$. This problem can be bypassed with the help of the scalar factor $\X$. Namely, there is a unique rescaling of $e^\a_\m$ by a function of $\X$ (up to an overall constant factor) that makes the torsion vanish. We thus define a proper coframe $\E^\a\equiv\E^\a_\m\dd x^\m$ by
\begin{equation}
\E^\a_\m\equiv e^\a_\m\sqrt{1-\s\X^2}.
\label{sGal:coframe}
\end{equation}
This indeed satisfies $\D\E^\a=0$ as a consequence of the identities in eq.~\eqref{sGal:MCrem}. It can thus be used to construct a metric,
\begin{equation}
G_{\m\n}\equiv \eta_{\a\b}\E^\a_\m\E^\b_\n=\eta_{\m\n}-\s\de_\m\de\p\cdot\de\de_\n\p.
\label{sGal:metric}
\end{equation}
Upon conversion of frame indices to Lorentz indices, the spin connection $-\mc^{\a\b}_L$ translates to the Levi-Civita connection of this metric, and thus can be used to construct covariant derivatives of spacetime tensors. The metric~\eqref{sGal:metric} was first derived in ref.~\cite{Novotny2017a} by a geometric construction based on the embedding of the physical Minkowski spacetime as a brane in a flat spacetime of doubled dimension.

The remaining content of eq.~\eqref{sGal:MCrem} translates to the condition $\eta_{\b\g}\tilde\mc^{\a\b}_S\w\E^\g=0$, where
\begin{equation}
\tilde\mc^{\a\b}_S\equiv\mc^{\a\b}_S-\frac{\eta^{\a\b}}{1-\s\X^2}\D\X\equiv\tilde\mc^{\a\b}_{S\l}\dd x^\l.
\label{sGal:MCStilde}
\end{equation}
As a consequence, the rank-3 covariant tensor $\O_{\m\n\l}\equiv\E_{\a\m}\E_{\b\n}\tilde\mc^{\a\b}_{S\l}$ is fully symmetric in its three indices. We still have not used the first two lines of eq.~\eqref{sGal:MCequations}. These are however easy to interpret. Namely, the first line gives the curvature 2-form of the spin connection $-\mc^{\a\b}_L$ in terms of $\mc^{\a\b}_S$. The second line guarantees the vanishing of the exterior covariant derivative $\D\mc^{\a\b}_S$. Both identities are readily translated in terms of the spacetime tensor $\O_{\m\n\l}$. Instead of working out the straightforward details, we now give a recap of the geometric structure of the sGal theory.

The geometric ingredients as provided by the technique of nonlinear realizations boil down to the coframe $\E^\a$ and the associated metric $G_{\m\n}$, given in terms of the NG field $\p$ respectively by eq.~\eqref{sGal:Xdef} combined with eq.~\eqref{sGal:coframe}, and eq.~\eqref{sGal:metric}. The metric comes equipped with the corresponding Levi-Civita connection, defined by the Christoffel symbols
\begin{equation}
\Gamma_{\l\m\n}=-\sigma\de_\l\de\phi\cdot\de\de_\m\de_\n\phi.
\end{equation}
Finally, there is the fully symmetric tensor $\O_{\m\n\l}$, which satisfies the Codazzi equation, equivalent to $\D\mc^{\a\b}_S=\D\tilde\mc^{\a\b}_S=0$,
\begin{equation}
\nabla_\k\O_{\l\m\n}=\nabla_\l\O_{\k\m\n}.
\label{sGal:Codazzi}
\end{equation}
This makes the rank-4 tensor $\nabla_\k\O_{\l\m\n}$ fully symmetric. The Riemann curvature tensor is related to the square of $\O_{\m\n\l}$ via the Gauss equation,
\begin{equation}
R_{\k\l\m\n}=-\s G^{\s\t}(\O_{\s\k\m}\O_{\t\l\n}-\O_{\s\k\n}\O_{\t\l\m}).
\label{sGal:Gauss}
\end{equation}
The information about the scalar factor $\X$, descending from the $\mc^\a_K$ part of the MC form, is encoded via eq.~\eqref{sGal:MCStilde} in the trace of $\O_{\m\n\l}$.

Note that similarly to the DBI theory, the special Galileon symmetry, generated by $S_{\m\n}$, acts on the spacetime coordinates as a field-dependent diffeomorphism, $\delta x^\m=-2\sigma s^{\m\n}\de_\n\phi\equiv\zeta^\m$, where $s^{\m\n}$ is a traceless symmetric matrix of infinitesimal parameters. The transformation of the NG field $\phi$ itself is highly nontrivial,
\begin{equation}
\delta\phi=-\zeta^\m\de_\m\phi-s^{\m\n}(x_\m x_\n+\sigma\de_\m\phi\de_\n\phi)=s^{\m\n}(-x_\m x_\n+\sigma\de_\m\phi\de_\n\phi).
\end{equation}
On the other hand, $\Omega_{\m\n\l}$ transforms as an ordinary spacetime tensor under the induced coordinate diffeomorphism,
\begin{equation}
\delta\Omega_{\m\n\l}=-\zeta^\k\de_\k\Omega_{\m\n\l}-\de_\m\zeta^\k\Omega_{\k\n\l}-\de_\n\zeta^\k\Omega_{\m\k\l}-\de_\l\zeta^\k\Omega_{\m\n\k}.
\end{equation}


\subsection{Equations of motion}
\label{subsec:sGalEoM}

Once expressed in terms of the above-constructed geometric ingredients, the structure of the sGal theory is independent of the parameterization of these objects in terms of the NG field. It only relies on the MC equations~\eqref{sGal:MCequations}, which in turn descend directly from the symmetry of the sGal theory. At this point, we need input from the actual dynamics of the theory, and we would naturally like to express it in terms of the same covariant objects.

To that end, we start with the Lagrangian of the sGal theory,
\begin{equation}
\L_\text{sGal}=\sum_{n=0}^Dc_n\varepsilon^{\m_1\dotsb\m_n\l_{n+1}\dotsb\l_D}\varepsilon^{\n_1\dotsb\n_n}_{\phantom{\n_1\dotsb\n_n}\l_{n+1}\dotsb\l_D}\p(\de_{\m_1}\de_{\n_1}\p)\dotsb(\de_{\m_n}\de_{\n_n}\p),
\label{sGal:lagrangian}
\end{equation}
where $c_{2k}=0$ and
\begin{equation}
c_{2k+1}=\frac{\s^k}D\binom{D}{2k+1}\frac{c_1}{k+1}.
\label{sGal:cn}
\end{equation}
The coefficient $c_1$ can be chosen arbitrarily and fixes the overall normalization of the Lagrangian. The first step towards the derivation of the EoM is to evaluate the variation of the action with respect to the NG field $\p$,
\begin{equation}
\frac{\d S_\text{sGal}}{\d\p}=\sum_{n=0}^D(n+1)c_n\varepsilon^{\m_1\dotsb\m_D}\varepsilon^{\n_1\dotsb\n_D}\tilde\Xi_{\m_1\n_1}\dotsb\tilde\Xi_{\m_n\n_n}\eta_{\m_{n+1}\n_{n+1}}\dotsb \eta_{\m_D\n_D},
\end{equation}
where $\tilde\Xi_{\m\n}\equiv\de_\m\de_\n\p$ differs from the previously introduced matrix $\Xi$ just by lowering the first index. Using eq.~\eqref{sGal:cn}, we find that the $n=2k+1$ term of the sum is nothing but the order-$\tilde\Xi^{2k+1}$ part of
\begin{equation}
\frac{2c_1}{D\sqrt\s}\varepsilon^{\m_1\dotsb\m_D}\varepsilon^{\n_1\dotsb\n_D}(\eta+\sqrt\s\tilde\Xi)_{\m_1\n_1}\dotsb(\eta+\sqrt\s\tilde\Xi)_{\m_D\n_D}=\frac{2c_1}{\sqrt\s}(-1)^{D-1}(D-1)!\det(\un+\sqrt\s\Xi).
\end{equation}
Upon collecting all the odd-$n$ terms but dropping the even-$n$ terms, we find that the EoM of the sGal theory is equivalent to $\det(\un+\sqrt\s\Xi)-\det(\un-\sqrt\s\Xi)=0$~\cite{Preucil2019}. This is in turn equivalent to
\begin{equation}
\log\det\frac{\un+\sqrt\s\Xi}{\un-\sqrt\s\Xi}=\tr\log\frac{\un+\sqrt\s\Xi}{\un-\sqrt\s\Xi}=0\qquad\Leftrightarrow\qquad\X=0,
\label{sGal:EoMX}
\end{equation}
where in the last step, we used eq.~\eqref{sGal:Xdef}.

The rest of the argument follows a familiar pattern. The vanishing of $\X$ implies by eq.~\eqref{sGal:MCStilde} that the tensor $\O_{\m\n\l}$ is traceless on solutions of the EoM,
\begin{equation}
G^{\m\n}\O_{\m\n\l}=0.
\label{sGal:EoMtrace}
\end{equation}
When combined with the Codazzi equation~\eqref{sGal:Codazzi}, this gives a covariant conservation law
\begin{equation}
0=\nabla_\m(G^{\k\l}\O_{\k\l\n})=G^{\k\l}\nabla_\m\O_{\k\l\n}=G^{\k\l}\nabla_\k\O_{\m\l\n}=\nabla^\l\O_{\l\m\n}.
\label{sGal:EoMcovcons}
\end{equation}
Taking then the covariant divergence of the Codazzi equation~\eqref{sGal:Codazzi} and manipulating the result \`a la eqs.~\eqref{NLSM:EoMderivation} and~\eqref{DBI:EoMderivation}, we find
\begin{equation}
\nabla^2\O_{\m\n\l}=R^{\s\t}_{\phantom{\s\t}\m\t}\O_{\s\n\l}-R^{\s\phantom\m\t}_{\phantom\s\m\phantom\t\n}\O_{\s\t\l}-R^{\s\phantom\m\t}_{\phantom\s\m\phantom\t\l}\O_{\s\t\n},
\end{equation}
where we traded the commutator of covariant derivatives for the Riemann curvature tensor. Finally, using the Gauss equation~\eqref{sGal:Gauss}, we arrive at a second-order EoM for the sGal theory that features a contact quartic self-interaction and is manifestly symmetric under permutations of its three indices,
\begin{equation}
\boxed{\nabla^2\O_{\m\n\l}-\s\O^{\r\s\t}(\O_{\m\n\r}\O_{\l\s\t}+\O_{\n\l\r}\O_{\m\s\t}+\O_{\l\m\r}\O_{\n\s\t})+2\s\O_{\m\r}^{\phantom{\m\r}\s}\O_{\n\s}^{\phantom{\n\s}\t}\O_{\l\t}^{\phantom{\l\t}\r}=0.}
\label{sGal:EoMscalar}
\end{equation}

Just like in the case of the DBI theory, the EoM~\eqref{sGal:EoMscalar} for the symmetric tensor field $\O_{\m\n\l}$ has to be augmented with a corresponding equation for the composite gravity sector, defined by the metric $G_{\m\n}$~\eqref{sGal:metric}. This is obtained by taking the trace of the Riemann curvature tensor and using the Gauss equation~\eqref{sGal:Gauss},
\begin{equation}
\boxed{R_{\m\n}=\s\O_{\k\l\m}\O^{\k\l}_{\phantom{\k\l}\n}.}
\label{sGal:EoMgravity}
\end{equation}


\subsection{Scattering amplitudes}

Equations~\eqref{sGal:EoMscalar} and~\eqref{sGal:EoMgravity} together constitute a closed set of second-order partial differential equations for the dynamical variables $\O_{\m\n\l}$ and $G_{\m\n}$. Similarly to section~\ref{subsec:DBI:amplitudesfromEoM}, we will illustrate their use by computing the four-point amplitude of the sGal theory. In the harmonic gauge for the composite metric $G_{\m\n}$, the EoM~\eqref{sGal:EoMgravity} for the graviton field $h_{\m\n}$ reduces upon linearization to
\begin{equation}
\Box h_{\m\n}\approx-2\s\O_{\k\l\m}\O^{\k\l}_{\phantom{\k\l}\n}.
\label{sGal:Riccilinearized}
\end{equation}
Similarly, the relevant part of the EoM~\eqref{sGal:EoMscalar} for the tensor current $\O_{\m\n\l}$ reads
\begin{align}
\Box\O_{\m\n\l}\approx&-2\s\O_{\m\r}^{\phantom{\m\r}\s}\O_{\n\s}^{\phantom{\n\s}\t}\O_{\l\t}^{\phantom{\l\t}\r}\\
\notag
&+h^{\r\s}\de_\r\de_\s\O_{\m\n\l}+\bigl[(\de_\m h^\k_{\phantom\k\s}+\de_\s h^\k_{\phantom\k\m}-\de^\k h_{\m\s})\de^\s\O_{\k\n\l}+\text{cycl. perm. of }\m,\n,\l\bigr].
\end{align}
The first line eventually yields the contact contribution to the one-point function $\langle\O_{\m\n\l}\rangle_J$ at the desired order in the external source $J$ for the NG field $\p$, whereas the second line contains contributions due to graviton exchange. Upon substituting for $h_{\m\n}$ from eq.~\eqref{sGal:Riccilinearized}, we arrive at an explicit expression for $\O^{(3)}_{\m\n\l}$ in terms of $\O^{(1)}_{\m\n\l}$, cf.~the notation introduced in eq.~\eqref{YMS:iteration},
\begin{equation}
\begin{split}
\Box\O_{\m\n\l}\approx&-2\s\O_{\m\r}^{\phantom{\m\r}\s}\O_{\n\s}^{\phantom{\n\s}\t}\O_{\l\t}^{\phantom{\l\t}\r}-2\s(\de_\r\de_\s\O_{\m\n\l})\Box^{-1}(\O_{\a\b}^{\phantom{\a\b}\r}\O^{\a\b\s})\\
&-2\s\bigl\{(\de^\s\O_{\k\n\l})\Box^{-1}[\de_\m(\O_{\a\b\s}\O^{\a\b\k})+\de_\s(\O_{\a\b\m}\O^{\a\b\k})-\de^\k(\O_{\a\b\m}\O^{\a\b}_{\phantom{\a\b}\s})]\\
&\phantom{-2\s\{\}}+\text{cycl. perm. of }\m,\n,\l\bigr\}.
\end{split}\label{sGal:V4}
\end{equation}

To convert the one-point function to the corresponding scattering amplitude, we use the fact that $\O_{\m\n\l}=\de_\m\de_\n\de_\l\phi+\mathcal O(\phi^2)$. What we have to do is thus to cut off the external propagators, take the on-shell limit, and attach to the external legs the root and leaf polarization tensors
\begin{equation}
\varepsilon^{\m\n\l}_\text{root}(p)=\frac{\im q^\m q^\n q^\l}{(p\cdot q)^3},\qquad
\varepsilon^{\m\n\l}_\text{leaf}(p)=-\im p^\m p^\n p^\l.
\label{sGal:polarizationvectors}
\end{equation}
The final result for the four-point amplitude is then
\begin{equation}
V_4^\text{on-shell}=-\frac\s4s_{12}s_{13}s_{14},
\end{equation}
which is manifestly invariant under arbitrary permutations of the momenta $p^\m_{i}$, $i=1,\dotsc,4$ on the external legs.


\section{Geometric duality of scattering amplitudes}
\label{sec:flavorkinematics}

The presentation of the YMS theory, the NLSM, and the DBI and sGal theories in terms of their classical EoM makes it straightforward to construct maps connecting the individual theories. As mentioned in the introduction, we thus find two different duality sequences of theories. The first of these includes the YMS theory and the NLSM, whereby the choice of the symmetric coset space $G/H$ for the latter dictates the structure of the gauge and scalar sectors of the former. We work out the details in section~\ref{subsec:YMS-NLSM}. The second duality sequence includes the NLSM\textsubscript{g}, the multiflavor DBI theory and the sGal theory, with details offered in sections~\ref{subsec:NLSMg-DBI} and~\ref{subsec:DBI-sGal}.

As we illustrated in the previous sections, on-shell scattering amplitudes of the EFTs considered in this paper can be extracted by a diagrammatic construction based on the second-order EoM for the tensor currents listed in table~\ref{tab:bigpicture} and for the corresponding gauge sectors, combined with the LSZ reduction. Therefore, to establish a duality map between the scattering amplitudes of two theories, all one has to do is to match the second-order EoMs for the respective gauge and matter sectors. \rev{This identifies the one-point functions in the different theories in presence of an external source, which in turn guarantees the existence of a duality map for all tree-level scattering amplitudes. We exemplify the duality more explicitly in appendix~\ref{app:4pt_vertices} by listing the four-point correlators in all the theories considered in this paper, with external legs already dressed by polarization vectors (tensors) but otherwise still off-shell. This highlights the analogous analytic structure of the correlators across the different theories.}


\subsection{From YMS theory to NLSM}
\label{subsec:YMS-NLSM}

For the reader's convenience, we reproduce the second-order EoMs for the matter and gauge sectors of the YMS theory, copied verbatim from section~\ref{sec:YMS},
 \begin{equation}
 \Da^2\p^{Ia}+\lambda g_{JK}g_{\a\b}g^{ab}f^\a_{bd}f^\b_{ce}\p^{Ic}\p^{Jd}\p^{Ke}=0, \qquad \Da_\m F^{\a\m\n}+g_{IJ}g^{\a\b}g_{ac}f^c_{\b b}(\Da^\n\p^{Ia})\p^{Jb}=0.
 \end{equation}
To make the duality between the YMS theory and the NLSM manifest, we make the assumption already utilized in section~\ref{subsec:YMS:amplitudesfromEoM}, namely that Greek and Latin indices herein can be raised and lowered respectively with the metrics $g_{\a\b}$ and $g_{ab}$. In addition, we assume that the symbols $f_{ABC}$ with three subscripts are fully antisymmetric; this is the case for instance when they descend from a compact symmetry group $G$ and $g_{\a\b}\oplus g_{ab}$ defines a decomposition of its Cartan-Killing form. Finally, we choose the basis of $Q_\a,Q_a$ so that $g_{\a\b}=\d_{\a\b}$, $g_{ab}=\d_{ab}$, and the flavor basis of the YMS scalar fields $\phi^{Ia}$ so that $g_{IJ}=\d_{IJ}$. This allows us to rewrite the equations for the YMS theory as
\begin{equation}
\Da^2\p^{Ia} + \lambda f^a_{\a b}f^\a_{cd}\p^{Jb}\p_J^{c} \p^{Id}=0,\qquad
\Da_\m F^{\a\m\n}- f^\a_{ab}(\Da^\n\p^{Ia})\p_I^{b}=0,
\label{YMS:EoMdeltaIJ}
\end{equation}
where we relabeled the color indices in the EoM for $\p^{Ia}$ to match the index structure of the EoM for $\mc^a_\m$ on the NLSM side. These can now be compared to the EoMs for the matter and gauge sector of the NLSM, again copied verbatim from section~\ref{sec:NLSM},
 \begin{align}
  \Da^2\mc^a_\m+f^a_{\a b}f^\a_{cd}\mc^b_\n\mc^{c\n}\mc^d_\m=0, \qquad \Da_\m F^{\a\m\n}+f^\a_{ab}(\Da^\n\mc^a_\m)\mc^{b\m}=0.
 \end{align}
It is clear that the equations for the NLSM can be recovered from those for the YMS theory if one fixes $\lambda=1$ and replaces
\begin{equation}
\p^{Ia}\to\mc^{a \m},\qquad
A^\a_\m\to-A^\a_\m,\qquad
\d_{IJ}\to \eta_{\m\n}.
\label{duality:YMS-NLSM}
\end{equation}
See table~\ref{tab:YMS-NLSM} for an overview. Up to the sign mismatch of the gauge fields (which is a consequence of the conventional definition of the MC form in the NLSM), the correspondence between the YMS theory and the NLSM amounts to a simple substitution of the dynamical variables and a replacement of flavor indices with Lorentz indices.

\begin{table}
\begin{center}
\makebox[\textwidth][c]{\begin{tabular}{c||c|c}
 & Matter sector & Gauge sector \\
\hline\hline
YMS & $\Da^2\p^{Ia} + f^a_{\a b}f^\a_{cd}\p^{Jb}\p_J^{c} \p^{Id}=0$ & 
 $\Da_\m F^{\a\m\n}- f^\a_{ab}(\Da^\n\p^{Ia})\p_I^{b}=0$ \\
\hline
NLSM & $\Da^2\mc^{a\m} + f^a_{\a b}f^\a_{cd}\mc^{b\n} \mc^c_\n \mc^{d\m}=0$ & $\Da_\m F^{\a\m\n}+f^\a_{ab}(\Da^\n \mc^{a\m})\mc^{b}_{\m}=0$
\end{tabular}}
\end{center}
\caption{\emph{The covariant EoMs for the matter and gauge sectors of the YMS-NLSM duality sequence, which map onto each other under the replacement~\eqref{duality:YMS-NLSM}. In the YMS theory, the matter and gauge degrees of freedom are respectively $\p^{Ia}$ and $A^\a_\m$. Their counterparts in the NLSM are $\mc^a_\m$ and the composite connection $A^\a_\m$. Note that we have set $\lambda =1$ on the YMS side.}}
\label{tab:YMS-NLSM}
\end{table}

This duality is manifested by the Feynman rules listed in figures~\ref{fig:YMSvertices} and~\ref{fig:NLSMvertices}. As such, it therefore trivially extends to the one-point function $\langle\p^{Ia}\rangle_J$ in the YMS theory and its counterpart $\langle\mc^a_\m\rangle_J$ in the NLSM. By the same token, the duality relates the on-shell scattering amplitudes in the two theories. Here, we however have to keep in mind that the prescriptions for extracting amplitudes from the one-point function in the two theories differ. Namely, in the NLSM, the one-point function stripped of the external source and the propagators on the external legs has to be decorated with the polarization vectors~\eqref{NLSM:polarizationvectors}. This shows in the relation between the (still off-shell) four-point correlators~\eqref{YMS:V4resultoffshell} and~\eqref{NLSM:V4resultoffshell}. Given that all the flavor indices in eq.~\eqref{YMS:V4resultoffshell} correspond to external particles, the NLSM correlator~\eqref{NLSM:V4resultoffshell} can be recovered therefrom by replacing the Kronecker delta factors with the products of the corresponding polarization vectors, $\d_{IJ}\to\varepsilon(p_I)\cdot\varepsilon(p_J)$ etc. The opposite overall sign is conventional and comes from the $\im$-factors on the polarization vectors.


\subsection{From NLSM\textsubscript{g} to DBI theory}
\label{subsec:NLSMg-DBI}

Writing the EoMs of the YMS theory in the form~\eqref{YMS:EoMdeltaIJ} makes it clear that the flavor symmetry of the theory is effectively $\gr{O}(n)$, where $n$ is the number of flavors. Upon mapping the YMS theory to the NLSM, this internal flavor symmetry is mapped to the spacetime Lorentz symmetry, and flavor indices are replaced with Lorentz indices. We end up with an EFT that has no flavor whatsoever, but the dynamical variable $\mc^a_\m$ in terms of which we express the classical EoM is a Lorentz vector.

In this and the next subsection, we will apply the same reasoning to the sequence NLSM\textsubscript{g}-DBI-sGal of theories. The final element of the sequence, that is the sGal theory, has no flavor and our matter field $\O_{\m\n\l}$ is a rank-3 Lorentz tensor. In contrast, the matter field of the DBI theory, $\K^I_{\m\n}$, is a rank-2 Lorentz tensor and carries a flavor index. This means that we have to start with a NLSM\textsubscript{g} whose dynamical variables carry \emph{two} flavor indices. To that end, consider a pseudo-Euclidean space of dimension $n+m$, carrying the action of generators $Q_{ij}$ satisfying the standard commutation relation,
\begin{equation}
[Q_{ij},Q_{kl}]=\im(g_{ik}Q_{jl}+g_{jl}Q_{ik}-g_{il}Q_{jk}-g_{jk}Q_{il}),
\label{Grassmann_commutator}
\end{equation}
where $g_{ij}$ is the pseudo-Euclidean metric with the convention that positive signature corresponds to spacelike geometry.

We now split the space into subspaces of dimensions $n$ and $m$, respectively. We will use capital Latin indices $I,J,\dotsc$ for the former, and barred capital indices $\bar I,\bar J,\dotsc$ for the latter. The metric thus splits as $g_{ij}=g_{IJ}\oplus g_{\bar I\bar J}$. In order to be able to construct a NLSM\textsubscript{g}-type Lagrangian with a positive-definite kinetic term, and at the same time match the algebraic structure of the DBI theory, we assume that both subspaces are strictly Euclidean, and together carry a linear representation of $\gr{O}(n)\times\gr{O}(m)$. This requires that each of the metrics $g_{IJ},g_{\bar I\bar J}$ is either positive- or negative-definite. Depending on the relative sign of the metrics, the Lie algebra defined by eq.~\eqref{Grassmann_commutator} then corresponds either to $\gr{SO}(n+m)$ or to $\gr{SO}(n,m)$.

We have thus established that a suitable starting point for studying the duality between the DBI and sGal theories is a NLSM\textsubscript{g} based on the coset space $\gr{SO}(n+m)/[\gr{SO}(n)\times\gr{SO}(m)]$ or $\gr{SO}(n,m)/[\gr{SO}(n)\times\gr{SO}(m)]$, which are both examples of a real Grassmannian. The unbroken generators constitute the pair $Q_{IJ},Q_{\bar I\bar J}$ of antisymmetric tensors, whereas the broken generators can be identified with $Q_{I\bar I}$. We can now reuse all the relations for the generic NLSM\textsubscript{g}, derived in section~\ref{subsec:NLSMg}. All we have to do is to use, wherever appropriate, the structure constant $f^{mn}_{ij,kl}$, defined by $[Q_{ij},Q_{kl}]\equiv(\im/2)f^{mn}_{ij,kl}Q_{mn}$ and found using eq.~\eqref{Grassmann_commutator},
\begin{equation}
f^{mn}_{ij,kl}=\d^m_i\d^n_kg_{jl}+\d^m_j\d^n_lg_{ik}-\d^m_i\d^n_lg_{jk}-\d^m_j\d^n_kg_{il}-(m\leftrightarrow n).
\end{equation}
The unbroken $\gr{SO}(n)\times\gr{SO}(m)$ symmetry forces the matrix coupling constant $\k_{ab}\to\k_{I\bar I,J\bar J}$ to factorize as $\k_{I\bar I,J\bar J}=\k g_{IJ}g_{\bar I\bar J}$; the remaining parameter $\k$ can be used to tune the strength of the coupling between the NLSM and gravity. Thus, the EoM~\eqref{NLSMg:EoMgravity} for the gravitational sector becomes
\begin{equation}
R_{\m\n}=-8\pi G\k g_{IJ}g_{\bar I\bar J}\mc^{I\bar I}_\m\mc^{J\bar J}_\n.
\label{Grassmann:EoMgravity}
\end{equation}

The doubly covariant derivative~\eqref{NLSMgDD} takes the form
\begin{equation}
\DD_\m\mc^{I\bar I}_\n=\nabla_\m\mc^{I\bar I}_\n+g_{JK}A^{IJ}_\m\mc^{K\bar I}_\n+g_{\bar J\bar K}A^{\bar I\bar J}_\m\mc^{I\bar K}_\n.
\label{Grassmann:covder}
\end{equation}
This fixes the structure of the second-order EoM~\eqref{NLSMg:EoMscalar} for the matter sector,
\begin{equation}
\begin{split}
\DD^2\mc^{I\bar I}_\m&+8\pi G\k g_{JK}g_{\bar J\bar K}\mc^{I\bar I}_\n\mc^{J\bar J\n}\mc^{K\bar K}_\m\\
&+g_{JK}g_{\bar J\bar K}(\mc^{K\bar K\n}\mc^{I\bar J}_\n\mc^{J\bar I}_\m+\mc^{K\bar K\n}\mc^{J\bar I}_\n\mc^{I\bar J}_\m-2\mc^{I\bar J}_\n\mc^{J\bar I\n}\mc^{K\bar K}_\m)=0.
\end{split}
\label{Grassmann:EoMscalar}
\end{equation}
In order to relate this equation to the EoMs of the other EFTs, it will be convenient to rearrange it as
\begin{equation}
    \DD^2\mc^{I\bar I}_{\m} + \mathcal{P}^I{}_{JKL, \vphantom{\bar{J}\bar{K}\bar{L}}}{}^{\bar I}{}_{\bar{J}\bar{K}\bar{L}} \mc^{J\bar J\n} \mc^{K\bar K}_{ \n} \mc^{L\bar L}_{ \m} = 0
\label{NLSM-g-EOM}
\end{equation}
in terms of the tensor
 \begin{equation}
     \begin{split}
         \mathcal{P}^I{}_{JKL, \vphantom{\bar{J}\bar{K}\bar{L}}}{}^{\bar I}{}_{\bar{J}\bar{K}\bar{L}}\equiv  &-\underbrace{ \d^I_J g_{KL} (\d^{\bar I}_{\bar K} g_{\bar J \bar L} - \d^{\bar I}_{\bar L} g_{\bar J \bar K})}_{\gr{SO}(m)}- \underbrace{ (\d^I_K g_{JL} - \d^I_L g_{JK}) \d^{\bar I}_{\bar J} g_{\bar K \bar L} }_{\gr{SO}(n)} \\
         &+\underbrace{8\pi G \kappa \d^I_J g_{KL} \d^{\bar I}_{\bar J} g_{\bar K \bar L}}_{\text{gravity}}.
     \end{split}
 \end{equation}
This tensor plays a similar role as the product of two structure constants in the EoMs for the YMS theory and the NLSM.\footnote{The tensor $\mathcal{P}$, however, has different symmetry properties than the product of two structure constants, and does not satisfy the Jacobi identity. It would be interesting to see whether this structure, due to the inclusion of gravity, has a different group-theoretic interpretation.} It consists of different parts, describing contact interactions for gravity (proportional to $G$) and a (composite) Yang-Mills sector. The co\-va\-riant derivatives in eq.~\eqref{NLSM-g-EOM} lead to accompanying exchange interaction for both~parts.

As to the gauge sector, we add for the record an expression for the field strength of the $\gr{SO}(n)$-connection $A^{IJ}_\m$,
\begin{equation}
\tilde{R}^{IJ}_{\phantom{IJ}\m\n}=\de_\m A^{IJ}_\n-\de_\n A^{IJ}_\m-g_{KL}(A^{IK}_\m A^{JL}_\n-A^{IK}_\n A^{JL}_\m).
\end{equation}
(For later convenience, we changed the symbol $F$ for the field strength, used in section~\ref{sec:NLSM}, to $\tilde{R}$.) An analogous expression applies to the field strength $\tilde{R}^{\bar I\bar J}_{\phantom{\bar I\bar J}\m\n}$ of the $\gr{SO}(m)$-connection $A^{\bar I\bar J}_\m$. As these are composite gauge fields, their field strengths are subject to the two analogs of the first identity in eq.~\eqref{NLSM:geomids},
\begin{equation}
        \tilde{R}^{IJ}{}_{\m\nu}  = g_{\bar I \bar J}\bigl(\mc^{I \bar I}_\m \mc^{J \bar J}_\n -\mc^{I \bar I}_\n \mc^{J \bar J}_\m \bigr) 
        \quad \mathrm{and} \quad    
        \tilde{R}^{\bar I \bar J}{}_{\m\nu}  = g_{ I  J}\bigl(\mc^{I \bar I}_\m \mc^{J \bar J}_\n -\mc^{I \bar I}_\n \mc^{J \bar J}_\m \bigr).
\end{equation}
Acting with an extra covariant derivative gives second-order EoMs that are special cases of eq.~\eqref{NLSMg:EoMgauge} in case of the Grassmannian coset space,
\begin{equation}
\DD_\m \tilde{R}^{IJ\m\n}+g_{\bar I\bar J}\bigl[(\DD^\n\mc^{I\bar I}_\m)\mc^{J\bar J\m}-(\DD^\n\mc^{J\bar J\m})\mc^{I\bar I}_\m\bigr]=0,
\label{Grassmann:EoMgauge}
\end{equation}
and
\begin{equation}
\DD_\m \tilde{R}^{\bar I\bar J\m\n}+g_{IJ}\bigl[(\DD^\n\mc^{I\bar I}_\m)\mc^{J\bar J\m}-(\DD^\n\mc^{J\bar J\m})\mc^{I\bar I}_\m\bigr]=0.
\label{Grassmann:EoMgaugebar}
\end{equation}

We have reached the point where we can set up a map from the NLSM\textsubscript{g} on the Grassmannian to the multiflavor DBI theory. This is defined, roughly speaking, by replacing barred flavor indices with frame indices of the DBI theory.\footnote{In section~\ref{subsec:YMS-NLSM}, we could replace flavor directly with Lorentz indices thanks to the fact that both theories therein were defined on a flat Minkowski spacetime. Here, the theories on both sides of the duality map are coupled to gravity. Using frame indices instead of Lorentz indices is required so that the constant metric $g_{\bar I\bar J}$ on the NLSM\textsubscript{g} side can be mapped to the constant metric $\eta_{\a\b}$ on the DBI side.} Concretely, we map
\begin{equation}
\mc^{I\bar I}_\m\to\K^{\a I}_\m,\qquad
A^{IJ}_\m\to\mc^{IJ}_{Q\m},\qquad
A^{\bar I\bar J}_\m\to\mc^{\a\b}_{L\m},\qquad
g_{\bar I\bar J}\to-\eta_{\a\b},
\label{duality:NLSMg-DBI}
\end{equation}
where we used the symbol $\K^{\a I}_\m$ instead of the $\mc^{\a I}_{K\m}$ introduced in section~\ref{sec:DBI} to avoid proliferation of indices. The tensor $\K^{\a I}_\m$ carries the same information as the extrinsic curvature tensor, $\K^I_{\m\n}\equiv \eta_{\a\b}e^\a_\m\K^{\b I}_\n$; cf.~eq.~\eqref{DBI:Kmunu}. In addition, the dynamical metric $g_{\m\n}$ of the NLSM\textsubscript{g} has to be mapped on the composite metric $G_{\m\n}$ of the DBI theory, and analogously for the corresponding Levi-Civita connections.

Upon the replacement~\eqref{duality:NLSMg-DBI}, the EoM~\eqref{Grassmann:EoMgravity} for the gravitational sector becomes $R_{\m\n}= 8\pi G \k g_{IJ}\eta_{\a\b}\K^{\a I}_\m\K^{\b J}_\n$. Choosing the scale of the matter sector as\footnote{The manifest positivity of $\k$ as implied by this relation seems to be in tension with the relation $\k_{I\bar I,J\bar J}=\k g_{IJ}g_{\bar I\bar J}$. Namely, the latter suggests that $\k$ be equal to the relative signature of the flavor metrics $g_{IJ}$ and $g_{\bar I\bar J}$ so as to ensure overall positivity of the kinetic term of the NLSM\textsubscript{g} on the Grassmannian manifold. Since we are only interested in perturbative duality of scattering amplitudes based on classical EoMs, we ignore issues such as this, arising from physical positivity bounds on energy.} 
\begin{equation}
\k=\frac1{8\pi G} ,
\label{scaleNLSM}
\end{equation}
then gives
\begin{equation}
R_{\m\n}=g_{IJ}\eta_{\a\b}\K^{\a I}_\m\K^{\b J}_\n,
\end{equation}
which matches the corresponding eq.~\eqref{DBI:EoMgravity} in the DBI theory, expressed in terms of $\K^{\a I}_\m$ instead of $\K^I_{\m\n}$. The covariant derivative~\eqref{Grassmann:covder} of the NLSM\textsubscript{g} on the Grassmannian is turned by the replacement~\eqref{duality:NLSMg-DBI} to
\begin{equation}
\DD_\m\K^{\a I}_\n=\nabla_\m\K^{\a I}_\n-\eta_{\b\g}\mc^{\a\b}_{L\m}\K^{\g I}_\n+g_{JK}\mc^{IJ}_{Q\m}\K^{\a K}_\n,
\label{DBI:covder2}
\end{equation}
where the covariant derivative $\nabla_\m$ only acts on the Lorentz index $\n$ of $\K^{\a I}_\n$. Altogether, the covariant derivative $\DD_\m\K^{\a I}_\n$ is equivalent to eq.~\eqref{DBI:covder} as it should. The matter EoM~\eqref{NLSM-g-EOM} of the NLSM\textsubscript{g} on the Grassmannian is then mapped to
\begin{equation}
\begin{split}
\DD^2 \K^{\a I}_{\m} + \mathcal{P}^I{}_{JKL, \vphantom{\bar{J}\bar{K}\bar{L}}}{}^{\a}{}_{\b \g \d} \K^{\b J\n} \K^{\g K }_{\n} \K^{\d L}_{\m} =0,
\end{split}\label{eq:DBIwithP}
\end{equation}
where we used the fixed scale~\eqref{scaleNLSM} and defined the tensor
\begin{equation}
\mathcal{P}^I{}_{JKL, \vphantom{\bar{J}\bar{K}\bar{L}}}{}^{\a}{}_{\b \g \d} \equiv \left. \mathcal{P}^I{}_{JKL, \vphantom{\bar{J}\bar{K}\bar{L}}}{}^{\bar I}{}_{\bar{J}\bar{K}\bar{L}}\right|_{g_{\bar I \bar J} \to -\eta_{\a \b},\kappa \to \tfrac{1}{8\pi G}}.
\label{Ptensor}
\end{equation}
This second-order EoM matches eq.~\eqref{DBI:EoMscalar} upon an appropriate conversion of the frame index on $\K^{\a I}_\m$ to a second Lorentz index. Next, the EoM~\eqref{Grassmann:EoMgauge} is mapped by the replacement~\eqref{duality:NLSMg-DBI} to
\begin{equation}
\DD_\m \tilde{R}^{IJ\m\n}-\eta_{\a\b}\bigl[(\DD^\n\K^{\a I}_\m)\K^{\b J\m}-(\DD^\n\K^{\b J\m})\K^{\a I}_\m\bigr]=0,
\end{equation}
which matches eq.~\eqref{DBI:EoMgauge}. Finally, the EoM~\eqref{Grassmann:EoMgaugebar} for the connection $\mc^{\bar I\bar J}_\m$ of the NLSM\textsubscript{g} on the Grassmann manifold is mapped to
\begin{equation}
\DD_\m R^{\a\b\m\n}-g_{IJ}\bigl[(\DD^\n\K^{\a I}_\m)\K^{\b J\m}-(\DD^\n\K^{\b J\m})\K^{\a I}_\m\bigr]=0,
\end{equation}
where the change of overall sign of the second term stems from the interpretation of $\mc^{\a\b}_{L\m}$ as minus the spin connection associated with the metric $G_{\m\n}$. This does not match any of the fundamental second-order equations of the DBI theory, it is nevertheless still correct. Namely, it descends from the Gauss equation~\eqref{DBI:Gauss} for the Riemann curvature tensor upon using the covariant conservation law for $\K^I_{\m\n}$ and the Codazzi equation~\eqref{DBI:Codazzi}.

\begin{table}
\begin{center}\makebox[\textwidth][c]{
\begin{tabular}{c||c|c}
Theory & Matter sector & Gauge sector \\
\hline\hline
NLSM\textsubscript{g} & $ \DD^2\mc^{I\bar I}_{\m} + \mathcal{P}^I{}_{JKL, \vphantom{\bar{J}\bar{K}\bar{L}}}{}^{\bar I}{}_{\bar{J}\bar{K}\bar{L}} \, \mc^{J\bar J\n} \mc^{K\bar K}_{ \n} \mc^{L\bar L}_{ \m} = 0$ & 
\rule[-5.5ex]{0pt}{12ex} $\begin{aligned}
R_{\m\n}&=-g_{IJ}g_{\bar I\bar J}\mc^{I\bar I}_\m\mc^{J\bar J}_\n\\
\tilde{R}^{\bar I \bar J}{}_{\m\nu} & = g_{ I  J}\bigl(\mc^{I \bar I}_\m \mc^{J \bar J}_\n -\mc^{I \bar I}_\n \mc^{J \bar J}_\m \bigr)\\
\tilde{R}^{IJ}{}_{\m\nu} & = g_{\bar I \bar J}\bigl(\mc^{I \bar I}_\m \mc^{J \bar J}_\n -\mc^{I \bar I}_\n \mc^{J \bar J}_\m \bigr)
\end{aligned}$\\
\hline
DBI & $\DD^2 \K^{\a I}_{\m} + \mathcal{P}^I{}_{JKL, \vphantom{\bar{J}\bar{K}\bar{L}}}{}^{\a}{}_{\b \g \d} \, \K^{\b J\n} \K^{\g K}_{\n} \K^{\d L}_{\m} =0$ &  
\rule[-3.5ex]{0pt}{8ex}
$\begin{aligned}
R^{\a \b}{}_{\mu \nu} &= -g_{IJ} \bigl( \K^{\a I}_\m \K^{\b J}_\n - \K^{\a I}_\n \K^{\b J}_\m \bigr)\\
\tilde{R}^{IJ}{}_{\m \n} &= -\eta_{\a\b} \bigl( \K^{\a I}_\m \K^{\b J}_\n - \K^{\a I}_\n \K^{\b J}_\m \bigr) 
\end{aligned}$\\ 
\hline
sGal & $\DD^2 \Omega_{\mu}^{\a \a{'}} + \mathcal{P}^{\a{'}}{}_{\b{'} \g{'} \d{'}, \vphantom{\bar{J}\bar{K}\bar{L}}}{}^{\a}{}_{\b \g \d} \, \Omega^{\b \b{'} \n} \Omega_{\nu}^{\g \g{'}  } \Omega_{\mu}^{\d \d{'} } =0$ & 
$\begin{aligned}
R^{\a \b}{}_{\mu \nu} = -\sigma \eta_{\g\d} \bigl( \O^{ \g \a}_\m \O^{\d \b}_\n - \O^{\g \a}_\n \O^{\d \b}_\m \bigr)
\end{aligned}$ \\ 
\end{tabular}}
\end{center}
\caption{\emph{The covariant EoMs for the matter and gauge sectors of the NLSM\textsubscript{g}-DBI-sGal duality sequence, which map onto each other under the replacements \eqref{duality:NLSMg-DBI} and \eqref{duality:DBI-sGal}. In the NLSM\textsubscript{g}, the matter degrees of freedom are $\mc^{I \bar I}_\m$ while the gauge sector consists of the composite connections $A^{IJ}_\m, A^{\bar I \bar J}_\m$ as well as dynamical gravity. Their counterparts in the DBI theory are $\K^{\a I}_{\m}$ and $\mc^{IJ}_{Q\m}$ plus the composite metric. Finally, in the sGal theory one has $\Omega_{\mu}^{\a \a{'}}$ plus the composite metric. Under the duality mappings, the gauge field strengths $\tilde R$ are converted into the Riemann tensor $R$. Note that we have set $\kappa=1/(8\pi G)$ in the NLSM\textsubscript{g}.}}
\label{tab:NLSMg-DBI-sGal}
\end{table}

Altogether, we have shown that upon the replacement~\eqref{duality:NLSMg-DBI}, all the dynamical equations of the NLSM\textsubscript{g} on the Grassmannian map to their appropriate counterparts in the multiflavor DBI theory, as also summarized in table~\ref{tab:NLSMg-DBI-sGal}. The one subtle feature of the duality map is that in the process, the broken part of the MC form $\mc^{I\bar I}_\m$ is mapped to the mixed tensor $\K^{\a I}_\m$, whereas all the dynamical equations of the DBI theory in section~\ref{sec:DBI} were expressed in terms of the purely spacetime symmetric tensor $\K^I_{\m\n}$. Luckily, $\K^{\a I}_\m$ and $\K^I_{\m\n}$ agree at the first order of their expansion in the NG fields $\p^I$. Hence the one-point functions $\langle\K^{\a I}_\m\rangle_J$ and $\langle\K^I_{\m\n}\rangle_J$ in presence of a background source for the NG fields are guaranteed to give the same on-shell scattering amplitudes upon LSZ reduction. However, off-shell, the two one-point functions will have somewhat different diagrammatic representations, since contracting the Lorentz indices on $\K^I_{\m\n}$ with the dynamical metric $G_{\m\n}$ instead of the constant frame metric $\eta_{\a\b}$ will lead to additional interaction vertices. \rev{In appendix~\ref{app:4pt_vertices}, we illustrate the duality explicitly by working out the details of four-point correlators in the NLSM\textsubscript{g} and the DBI theory, based on the $\mc^{I\bar I}_\m$ and $\K^{\a I}_\m$ variables. This makes the duality manifest at the level of off-shell correlators.}


\subsection{From DBI theory to sGal theory}
\label{subsec:DBI-sGal}

The final step from the DBI theory to the sGal theory follows the same philosophy as the map from the NLSM\textsubscript{g} to the DBI theory, we will therefore be more concise. What we have to do is to map the remaining flavor index of the DBI theory to a new frame index. More precisely, we have to apply the replacement rules
\begin{equation}
\K^I_{\m\n}\to\O^\a_{\m\n},\qquad
\mc^{IJ}_{Q\m}\to-\s\mc^{\a\b}_{L\m},\qquad
g_{IJ}\to\s \eta_{\a\b},
\label{duality:DBI-sGal}
\end{equation}
where the tensor $\O^\a_{\m\n}$ is related to the rank-3 symmetric tensor field of the sGal theory by converting one of the Lorentz indices into a frame index, $\O_{\l\m\n}\equiv \E_{\a\l}\O^\a_{\m\n}$. Finally, the composite metric $G_{\m\n}$ along with its Levi-Civita connection are mapped to the corresponding objects in the sGal theory.

The scheme for computation of scattering amplitudes of the DBI theory from classical EoMs is based on the three framed equations, listed in section~\ref{subsec:DBIEoM}. The last of these, that is the EoM~\eqref{DBI:EoMgravity} for the composite graviton, is obviously turned by the map~\eqref{duality:DBI-sGal} to the corresponding EoM~\eqref{sGal:EoMgravity} in the sGal theory. To make further progress, we note that eq.~\eqref{duality:DBI-sGal} turns the covariant derivative~\eqref{DBI:covder} of the DBI theory to
\begin{equation}
\DD_\m\O^\a_{\n\l}=\nabla_\m\O^\a_{\n\l}-\eta_{\b\g}\mc^{\a\b}_{L\m}\O^\g_{\n\l},
\end{equation}
where the covariant derivative $\nabla_\m$ only acts on the Lorentz indices $\n,\l$ of $\O^\a_{\n\l}$. This is equivalent to the covariant derivative $\nabla_\k\O_{\l\m\n}$ of the rank-3 tensor $\O_{\l\m\n}$ as it should. With this detail in place, it is straightforward to check that eq.~\eqref{DBI:EoMscalar} maps to the EoM~\eqref{sGal:EoMscalar} for the matter sector of the sGal theory.

This already covers the two basic second-order equations of the sGal theory as listed in section~\ref{subsec:sGalEoM}, that is
\begin{equation}
    \DD^2 \Omega_{\mu}^{\a{'} \a} + \mathcal{P}^{\a{'}}{}_{\b{'} \g{'} \d{'}, \vphantom{\bar{J}\bar{M}\bar{N}}}{}^{\a}{}_{\b \g \d} \Omega^{\b{'} \b \n} \Omega_{\nu}^{\g{'} \g  } \Omega_{\mu}^{\d{'} \d } =0\quad \text{and}\quad
    R_{\m\n}= \sigma\eta_{\a\b}\eta_{\g\d}\O^{\a \g}_\m \O^{\b \d}_\n,\label{eq:sGalwithP}
\end{equation}
where the analog of the tensor~\eqref{Ptensor} is now defined by
\begin{equation}
\mathcal{P}^{\a{'}}{}_{\b{'} \g{'} \d{'}, \vphantom{\bar{J}\bar{K}\bar{L}}}{}^{\a}{}_{\b \g \d}
\equiv\left.\mathcal{P}^I{}_{JKL, \vphantom{\bar{J}\bar{K}\bar{L}}}{}^{\a}{}_{\b \g \d} \right|_{g_{IJ} \to \sigma \eta_{\a \b}} = \left. \mathcal{P}^I{}_{JKL, \vphantom{\bar{J}\bar{K}\bar{L}}}{}^{\bar I}{}_{\bar{J}\bar{K}\bar{L}}\right|_{g_{IJ} \to \sigma\eta_{\a \b},g_{\bar I \bar J} \to -\eta_{\a \b},\kappa \to \tfrac{1}{8\pi G}}.
\label{triple-mapping}
\end{equation}
However, we still have to deal with eq.~\eqref{DBI:EoMgauge}, which in the DBI theory has the status of a separate EoM for the gauge connection of the $\gr{O}(n)$ flavor symmetry. This is mapped by the replacement rules~\eqref{duality:DBI-sGal} to
\begin{equation}
\DD_\m R^{\a\b\m\n}=\s\bigl[(\DD^\n\O^\a_{\m\l})\O^{\b\m\l}-(\DD^\n\O^\b_{\m\l})\O^{\a\m\l}\bigr],
\label{sGal:EoMaux}
\end{equation}
where we again used the fact that $\mc^{\a\b}_{L\m}$ is minus the spin connection associated with the composite metric $G_{\m\n}$. Equation~\eqref{sGal:EoMaux} for the frame components of the Riemann curvature tensor is not among the basic EoMs of the sGal theory, but it is still correct. Namely, it descends from the Gauss equation~\eqref{sGal:Gauss} upon using the covariant conservation law for $\O_{\m\n\l}$ along with the Codazzi equation~\eqref{sGal:Codazzi}. This completes the setup of the duality map between the classical EoMs of the DBI and sGal theories, again summarized in table~\ref{tab:NLSMg-DBI-sGal}. \rev{An explicit illustration of the duality between the DBI and sGal theories at the level of four-point off-shell correlators is provided in appendix~\ref{app:4pt_vertices}. In order to connect naturally to the duality between the NLSM\textsubscript{g} and the DBI theory, this illustration is based on the $\K^{\a I}_\m$ and $\Omega^{\a'\a}_\m$ field variables.}


\section{Conclusions and outlook}
\label{sec:conclusions}

The key observation put forward in this paper is that a collection of well-known theories --- the YMS theory, the NLSM (with or without gravity), the multiflavor DBI theory, and the sGal theory --- can be formulated in a manifestly gauge-covariant manner that features a single, contact four-point interaction vertex. We believe that this result is potentially valuable in and of itself. Indeed, in a companion paper~\cite{Li:2024bwq}, we show how our quartic reformulation of the NLSM can be utilized to manifest the vanishing of the scattering amplitudes in the NLSM on a certain kinematical locus, generalizing the Adler zero principle~\cite{ArkaniHamed2023}. It would be interesting to investigate whether similar considerations also apply to the covariant formulations of the DBI and sGal theories.

Perhaps even more suprprisingly, our formulation of the classical EoM reveals an extremely simple duality relationship between the said theories. This essentially relies on the geometric nature of the building blocks of the classical EoM. The correspondence between different theories is one between the geometry of spacetime and of the target space of the theory, or the intrinsic and extrinsic geometry thereof. As we have discussed, this naturally leads to two sequences of dual theories, YMS-NLSM and NLSM\textsubscript{g}-DBI-SG. The former sequence with the NLSM describes NG bosons due to spontaneously broken internal symmetry, while the latter sequence involves EFTs with spontaneously broken spacetime symmetries. Interestingly, all these EFTs are therefore related to propagating gluons (in the YMS theory) and gravitons (in the NLSM\textsubscript{g}) under the geometric duality mappings.

The flavor-kinematics duality outlined in this paper can be seen as complementary to color-kinematics duality, which was demonstrated at the level of EoM in ref.~\cite{Cheung2021}. The starting point in the latter approach is a cubic interaction for scalars that are adjoint-valued with respect to two color groups. Replacing one color group with kinematics leads to a formulation of the NLSM that also features a vector field (in this case the Noether current for the spontaneously broken chiral symmetry). Subsequent replacement of the second color group yields the sGal theory formulated in terms of a symmetric rank-2 tensor. The main difference with our approach lies in the valency of the central vertex, being cubic rather than quartic. As we have shown, the latter case requires in addition the inclusion of gauge and/or gravity sectors.

Our work opens several natural avenues for future research. First and foremost, realizing the duality between different theories at the level of their classical EoM raises the immediate question whether one can relate not only the scattering of asymptotic one-particle states, but also fully nonlinear, exact solutions of the EoM. This is indeed the case, although the application of the ideas developed here to exact classical solutions is somewhat delicate. We therefore defer details to a separate, forthcoming publication.

Second, the flavor-kinematics duality of the YMS-NLSM and NLSM\textsubscript{g}-DBI-sGal sequences is naturally embedded in a larger web of dualities between EFTs for NG bosons. Guided by the duality tetrahedron of ref.~\cite{Neeling2022}, we expect a similar relationship between the (multiflavor) Maxwell theory coupled to gravity and the Born-Infeld theory. In the future, we aim to make this duality explicit using the classical EoM approach developed here.

Last but not least, one great advantage of the classical EoM approach is that it does not rely on the specifics of relativistic kinematics, as long as the building blocks of the EoM can be given a geometric interpretation. This opens the window to applications of double copy techniques to nonrelativistic field theory, where even parity of interaction vertices may be required by charge conservation. Our preliminary work suggests that the EoM of EFTs for nonrelativistic (Schr\"odinger-like) NG bosons indeed can be cast in a covariant form with a contact quartic coupling along the same line as in the relativistic NLSM. While there is no nonrelativistic analog of the sGal theory, a nonrelativistic version of the DBI theory does exist~\cite{Brauner2021a} and features prominently in the landscape of exceptional nonrelativistic EFTs~\cite{Mojahed2022}.

We conclude by pointing out that ours is not the only work to emphasize the relevance of geometry for understanding the structure and detailed properties of scattering amplitudes in EFTs of NG bosons. Indeed, the geometry of field space has provided valuable tools to disentangle observable physical properties of such EFTs from the artifacts of their Lagrangian description; see e.g.~refs.~\cite{Cheung2022,Cheung2022c,Craig2023a}. It has also been utilized to establish double-copy maps between amplitudes in different theories~\cite{Helset2024}. We expect the importance of geometry for understanding of detailed properties of EFTs, and of relations between different EFTs, to continue growing in the future.


\acknowledgments

We thank Christoph Bartsch, Nichita Berzan, Zongzhe Du, Song He, Scott Melville, Jasper Roosmale Nepveu, David Stefanyszyn and Tonnis ter Veldhuis for helpful discussions and feedback on our results. T.W.~is supported by China Scholarship Council. 


\appendix

\section{\rev{Four-point off-shell correlators}}
\label{app:4pt_vertices}

\rev{In this appendix, we list four-point off-shell correlators in all the theories considered in this paper, and show how they are mapped to each other by duality. We start with the YMS-NLSM duality sequence of theories. Here, the four-point correlators are derived in detail in the main text, and we thus just repeat the results for the reader's convenience. The four-point off-shell correlator of the YMS theory is given by eq.~\eqref{YMS:V4resultoffshell},
\begin{equation}
\begin{split}
V_{4\text{YMS}}^\text{off-shell}={}&f_{\a ab}f^\a_{cd}\biggl[\lambda (\d_{IL}\d_{JK}-\d_{IK}\d_{LJ})+2\d_{IJ}\d_{KL}\frac{p_J\cdot(p_K-p_L)}{s_{IJ}}\biggr]\\
&+f_{\a ac}f^\a_{db}\biggl[\lambda(\d_{IJ}\d_{KL}-\d_{IL}\d_{JK})+2\d_{IK}\d_{LJ}\frac{p_K\cdot(p_L-p_J)}{s_{IK}}\biggr]\\
&+f_{\a ad}f^\a_{bc}\biggl[\lambda(\d_{IK}\d_{LJ}-\d_{IJ}\d_{KL})+2\d_{IL}\d_{JK}\frac{p_L\cdot(p_J-p_K)}{s_{IL}}\biggr],
\end{split}
\end{equation}
while that of the NLSM is given by eq.~\eqref{NLSM:V4resultoffshell},
\begin{equation}
\begin{split}
V_{4\text{NLSM}}^\text{off-shell}=&-f_{\a ab}f^\a_{cd}\biggl[\frac{q\cdot p_d}{q\cdot p_a}(p_b\cdot p_c)-\frac{q\cdot p_c}{q\cdot p_a}(p_d\cdot p_b)+\frac{q\cdot p_b}{q\cdot p_a}\frac{2}{s_{ab}}(p_c\cdot p_d)p_b\cdot(p_c-p_d)\biggr]\\
&-f_{\a ac}f^\a_{db}\biggl[\frac{q\cdot p_b}{q\cdot p_a}(p_c\cdot p_d)-\frac{q\cdot p_d}{q\cdot p_a}(p_b\cdot p_c)+\frac{q\cdot p_c}{q\cdot p_a}\frac{2}{s_{ac}}(p_d\cdot p_b)p_c\cdot(p_d-p_b)\biggr]\\
&-f_{\a ad}f^\a_{bc}\biggl[\frac{q\cdot p_c}{q\cdot p_a}(p_d\cdot p_b)-\frac{q\cdot p_b}{q\cdot p_a}(p_c\cdot p_d)+\frac{q\cdot p_d}{q\cdot p_a}\frac{2}{s_{ad}}(p_b\cdot p_c)p_d\cdot(p_b-p_c)\biggr].
\end{split}
\end{equation}
These four-point correlators can be identified by setting $\l=1$ in $V_{4\text{YMS}}^\text{off-shell}$, making the substitution $\delta_{IJ}\to \varepsilon(p_{I})\cdot\varepsilon(p_{J})$ which implements the polarization vectors attached to the external legs in the NLSM, and finally reducing the labels $Ia,Jb,Kc,Ld$ on the external legs of the correlator of the YMS theory to just $a,b,c,d$. This agrees with the duality map~\eqref{duality:YMS-NLSM} at the level of the EoM.}

\rev{To address the NLSM\textsubscript{g}-DBI-sGal duality sequence, we first sketch the derivation of the four-point correlator in the NLSM\textsubscript{g} on the Grassmannian, which was omitted in the main text. The first step is to expand the covariant derivatives in eqs.~\eqref{Grassmann:EoMgravity}, \eqref{Grassmann:EoMscalar}, \eqref{Grassmann:EoMgauge} and~\eqref{Grassmann:EoMgaugebar}. Upon imposing the harmonic gauge on the metric field and the Lorenz gauge on the gauge connections, the EoMs for the NLSM\textsubscript{g} take the form, to the order required for the computation of the four-point correlator,
\begin{equation}
\begin{split}
    \Box \mc^{I\bar I}_{\m} ={}& - \mathcal{P}^I{}_{JKL, \vphantom{\bar{J}\bar{K}\bar{L}}}{}^{\bar I}{}_{\bar{J}\bar{K}\bar{L}} \mc^{J\bar J\n} \mc^{K\bar K}_{ \n} \mc^{L\bar L}_{ \m}
    +h^{\n\k}\de_{\n}\de_{\k}\mc^{I\bar I}_{\m}\\
    &+\de^{\k}\Gamma^{\n}_{\m\k}\mc^{I\bar I}_{\n}+2\Gamma^{\n}_{\m\k}\de^{\k}\mc^{I\bar I}_{\n}
    -2g_{JK}A^{IJ}_{\n}\de^{\n}\mc^{K\bar I}_{\m}-2g_{\bar{J}\bar{K}}A^{\bar{I}\bar{J}}_{\n}\de^{\n}\mc^{I\bar K}_{\m}+\dotsb,\\
    \Box h_{\m\n}={}&16\pi G \k g_{IJ}g_{\bar I \bar J}\mc^{I\bar I}_{\m} \mc^{J \bar J}_{\n}+\dotsb,\\
    \Box A^{IJ}_{\m}={}&-g_{\bar I \bar J}\bigr(\de_{\m}\mc^{I\bar I}_{\n}\mc^{J\bar J \n}-\mc^{I\bar I}_{\n}\de_{\m}\mc^{J\bar J \n}\bigr)+\dotsb,\\
    \Box A^{\bar I\bar J}_{\m}={}&-g_{IJ}\bigr(\de_{\m}\mc^{I\bar I}_{\n}\mc^{J\bar J \n}-\mc^{I\bar I}_{\n}\de_{\m}\mc^{J\bar J \n}\bigr)+\dotsb.
\end{split} \label{eq:NLSMg_expansion}   
\end{equation}
The next step is to combine the above equations to produce a single expression for the one-point function of $\mc^{I\bar I}_\m$. Upon translating everything into momentum space and dressing the external legs with appropriate polarization vectors, this gives the four-point off-shell correlator,
\begin{equation}
    \begin{split}
        V_{4\text{NLSM\textsubscript{g}}}^\text{off-shell}={}&\frac{q\cdot p_J}{q\cdot p_I} \d_{IJ}g_{KL}
        \left(
        \begin{split}
            &\left(\d_{\bar I \bar K}g_{\bar J\bar L}+\d_{\bar I\bar L}g_{\bar J\bar K}\right)(p_K \cdot p_L)
            \\
            &-\left(\d_{\bar I \bar K}g_{\bar J\bar L}-\d_{\bar I\bar L}g_{\bar J\bar K}\right)\frac{2}{s_{KL}}(p_K \cdot p_L)(p_J \cdot (p_K-p_L))
        \end{split}
        \right)\\
        &+\frac{q\cdot p_K}{q\cdot p_I} \d_{IJ}g_{KL}
        \left(
        \begin{split}
            &\left(\d_{\bar I \bar K}g_{\bar J\bar L}-2\d_{\bar I\bar L}g_{\bar J\bar K}\right)(p_J \cdot p_L)
            \\
            &-\d_{\bar I\bar K}g_{\bar J\bar L}\frac{2}{s_{JL}}(p_J \cdot p_L)(p_K \cdot (p_J-p_L))
        \end{split}
        \right)\\
        &+\frac{q\cdot p_L}{q\cdot p_I} \d_{IJ}g_{KL}
        \left(
        \begin{split}
            &\left(-2\d_{\bar I \bar K}g_{\bar J\bar L}+\d_{\bar I\bar L}g_{\bar J\bar K}\right)(p_J \cdot p_K)
            \\
            &-\d_{\bar I\bar L}g_{\bar J\bar K}\frac{2}{s_{JK}}(p_J \cdot p_K)(p_L \cdot (p_J-p_K))
        \end{split}
        \right)\\
        &-\frac{q\cdot (p_J+p_K+p_L)}{q\cdot p_I} \d_{IJ}g_{KL}\d_{\bar I \bar J}g_{\bar K \bar L}\frac{32\pi G\k}{s_{KL}}(p_J \cdot p_K)(p_J \cdot p_L)+\text{cyclic},
    \end{split}
\end{equation}
where the same expression is to be summed over cyclic permutations of the pairs $\{\{J,\bar J\}$, $\{K,\bar K\}$, $\{L,\bar L\}\}$. In the on-shell limit, we arrive at the four-point amplitude of the NLSM\textsubscript{g},
\begin{equation}
V_{4\text{NLSM\textsubscript{g}}}^\text{on-shell}=s_{IK}\left(\d_{IJ}g_{KL}\d_{\bar I \bar L}g_{\bar J \bar K}+\d_{IL}g_{JK}\d_{\bar I\bar J}g_{\bar K \bar L}\right)+8\pi G \k \frac{s_{IK}s_{IL}}{s_{IJ}}\d_{IJ}g_{KL}\d_{\bar I\bar J}g_{\bar K \bar L}+\text{cyclic},
\end{equation}
which agrees with eq.~(A.9) of ref.~\cite{Neeling2022}, modulo the choice of conventions.}

\rev{To complete the discussion of the second duality sequence of EFTs, we also provide the four-point off-shell correlators for the remaining two theories. For the DBI theory, this is given by expanding the covariant derivatives in eq.~\eqref{eq:DBIwithP},
\begin{equation}
\begin{split}
    \Box \K^{I\a}_{\m} ={}& - \mathcal{P}^I{}_{JKL, \vphantom{\b\g\d}}{}^{\a}{}_{\b\g\d} \K^{J\b\n} \K^{K\g}_{ \n} \K^{L\d}_{ \m}
    +h^{\n\k}\de_{\n}\de_{\k}\K^{I\a}_{\m}\\
    &+\de^{\k}\Gamma^{\n}_{\m\k}\K^{I\a}_{\n}+2\Gamma^{\n}_{\m\k}\de^{\k}\K^{I\a}_{\n}-2g_{JK}A^{IJ}_{\n}\de^{\n}\K^{K\a}_{\m}+2\eta_{\b\g}\mc^{\a\b}_{L\n}\de^{\n}\K^{I\g}_{\m}+\dotsb,\\
    \Box h_{\m\n}={}&-2 g_{IJ}\eta_{\a\b}\K^{I\a}_{\m} \K^{J \b}_{\n}+\dotsb,\\
    \Box A^{IJ}_{\m}={}&\eta_{\a \b}\bigr(\de_{\m}\K^{I\a}_{\n}\K^{J\b \n}-\K^{I\a}_{\n}\de_{\m}\K^{J\b \n}\bigr)+\dotsb,\\
    \Box \mc^{\a\b}_{L\m}={}&-g_{IJ}\bigr(\de_{\m}\K^{I\a}_{\n}\K^{J\b \n}-\K^{I\a}_{\n}\de_{\m}\K^{J\b \n}\bigr)+\dotsb,
\end{split}    
\end{equation}
which precisely reproduces eq.~\eqref{eq:NLSMg_expansion} once the replacement rule~\eqref{duality:NLSMg-DBI} is enforced. Upon transforming to momentum space and dressing the external legs with root and leaf polarizations corresponding to the operator $\K^{I\a}_\m$, we arrive at the four-point off-shell correlator,
\begin{align}
\notag
    V_{4\text{DBI}}^\text{off-shell}=&-\frac{q\cdot p_J}{q\cdot p_I}\frac{q\cdot (p_J+2p_K+2p_L)}{q\cdot p_I}\d_{IJ}g_{KL}\frac{4}{s_{KL}}(p_J \cdot p_K)(p_J \cdot p_L)(p_K \cdot p_L)\\
    \notag
    &+\left(\frac{q\cdot p_K}{q\cdot p_I}\right)^2 \d_{IJ}g_{KL}(p_J \cdot p_L)^2\left(1-\frac{2p_K \cdot (p_J - p_L)}{s_{JL}}\right)\\
    &+\left(\frac{q\cdot p_L}{q\cdot p_I}\right)^2 \d_{IJ}g_{KL} (p_J \cdot p_K)^2\left(1-\frac{2p_L \cdot (p_J - p_K)}{s_{JK}}\right)\\
    \notag
    &-4\frac{q\cdot p_K}{q\cdot p_I}\frac{q\cdot p_L}{q\cdot p_I} \d_{IJ}g_{KL}(p_J \cdot p_K)(p_J \cdot p_L)\\
    \notag
    &+\left(\frac{q\cdot p_J}{q\cdot p_I}\right)\left(\frac{q\cdot p_K}{q\cdot p_I}(p_J \cdot p_L)+\frac{q\cdot p_L}{q\cdot p_I}(p_J \cdot p_K)\right)\\
    \notag
    &\qquad\quad\times \d_{IJ}g_{KL}  (p_K \cdot p_L) \left(1+\frac{2p_J\cdot(p_K+p_L)}{s_{KL}}\right)+\text{cyclic},
\end{align}
where ``cyclic'' now implies a sum over cyclic permutations of $\{J,K,L\}$.}

\rev{Finally, the four-point off-shell correlator of the sGal theory follows from eq.~\eqref{eq:sGalwithP} and reads
\begin{align}
    V_{4\text{sGal}}^\text{off-shell}={}&4\s\left(\frac{q\cdot p_J}{q\cdot p_I}\right)^2\frac{q\cdot (p_J+3p_K+3p_L)}{q\cdot p_I}\frac{p_K \cdot p_L}{s_{KL}}(p_J \cdot p_K)(p_J \cdot p_L)(p_K \cdot p_L)\notag\\
    &+4\s\frac{q\cdot p_J}{q\cdot p_I}\frac{q\cdot p_K}{q\cdot p_I}\frac{q\cdot p_L}{q\cdot p_I}(p_J \cdot p_K)(p_J \cdot p_L)(p_K \cdot p_L)\notag\\
    &-2\s\frac{q\cdot p_J}{q\cdot p_I}\left(\frac{q\cdot p_K}{q\cdot p_I}\right)^2 (p_J \cdot p_L)^2 (p_K \cdot p_L) \left(1+\frac{2p_K\cdot(p_J+p_L)}{s_{JL}}\right)\\
    &-2\s\left(\frac{q\cdot p_J}{q\cdot p_I}\right)^2\frac{q\cdot p_K}{q\cdot p_I} (p_K \cdot p_L)^2 (p_J \cdot p_L) \left(1+\frac{2p_J\cdot(p_K+p_L)}{s_{KL}}\right)+\text{cyclic}.\notag
\end{align}
The off-shell correlator of the NLSM\textsubscript{g} can now be mapped to that of the DBI theory and in turn to that of the sGal theory. The step from the NLSM\textsubscript{g} to the DBI theory requires one to set $\k=1/(8\pi G)$ and $g_{\bar I \bar J} = - \d_{\bar I\bar J}$, and subsequently replace $\d_{\bar I\bar J}\to\varepsilon(p_I)\cdot\varepsilon(p_J)$ and reduce the labels $I\bar I,J\bar J,K\bar K,L\bar L$ on the external legs to mere $I,J,K,L$. Finally, the step from the DBI to the sGal theory requires setting $g_{IJ}=\s\d_{IJ}$, replacing $\delta_{IJ}\to \varepsilon(p_I)\cdot \varepsilon(p_J)$, and eventually removing the flavor labels $I,J,K,L$ on the external legs. These substitution rules are in accord with the duality maps, discovered at the level of the EoM and given by eqs.~\eqref{duality:NLSMg-DBI} and~\eqref{duality:DBI-sGal}. Note that the successive replacements of the Kronecker $\d$s build up the rank-2 and rank-3 polarization tensors of the DBI and sGal theories, as given in eqs.~\eqref{DBI:polarizationvectors} and \eqref{sGal:polarizationvectors}: these are simply products of the NLSM polarization vectors~\eqref{NLSM:polarizationvectors}.}


\bibliographystyle{JHEP}
\bibliography{references}

\providecommand{\href}[2]{#2}\begingroup\raggedright\begin{thebibliography}{10}

\bibitem{Elvang2015}
H.~Elvang and Y.-t.~Huang, \emph{{Scattering Amplitudes in Gauge Theory and
  Gravity}}, Cambridge University Press, Cambridge, UK (2015).

\bibitem{Britto2005}
R.~Britto, F.~Cachazo and B.~Feng, \emph{{New recursion relations for tree
  amplitudes of gluons}},
  \href{https://doi.org/10.1016/j.nuclphysb.2005.02.030}{\emph{Nucl. Phys.}
  {\bfseries B715} (2005) 499}
  [\href{https://arxiv.org/abs/hep-th/0412308}{{\ttfamily hep-th/0412308}}].

\bibitem{Britto2005a}
R.~Britto, F.~Cachazo, B.~Feng and E.~Witten, \emph{{Direct Proof of the
  Tree-Level Scattering Amplitude Recursion Relation in Yang-Mills Theory}},
  \href{https://doi.org/10.1103/PhysRevLett.94.181602}{\emph{Phys. Rev. Lett.}
  {\bfseries 94} (2005) 181602}
  [\href{https://arxiv.org/abs/hep-th/0501052}{{\ttfamily hep-th/0501052}}].

\bibitem{Bedford2005}
J.~Bedford, A.~Brandhuber, B.J.~Spence and G.~Travaglini, \emph{{A Recursion
  relation for gravity amplitudes}},
  \href{https://doi.org/10.1016/j.nuclphysb.2005.016}{\emph{Nucl. Phys.}
  {\bfseries B721} (2005) 98}
  [\href{https://arxiv.org/abs/hep-th/0502146}{{\ttfamily hep-th/0502146}}].

\bibitem{Cachazo2005}
F.~Cachazo and P.~{Svr\v{c}ek}, \emph{{Tree level recursion relations in
  general relativity}},  \href{https://arxiv.org/abs/hep-th/0502160}{{\ttfamily
  hep-th/0502160}}.

\bibitem{Kampf2013a}
K.~Kampf, J.~{Novotn\'y} and J.~Trnka, \emph{{Recursion relations for
  tree-level amplitudes in the $SU(N)$ nonlinear sigma model}},
  \href{https://doi.org/10.1103/PhysRevD.87.081701}{\emph{Phys. Rev.}
  {\bfseries D87} (2013) 081701}
  [\href{https://arxiv.org/abs/1212.5224}{{\ttfamily 1212.5224}}].

\bibitem{Cheung2016a}
C.~Cheung, K.~Kampf, J.~{Novotn\'y}, C.-H.~Shen and J.~Trnka, \emph{{On-Shell
  Recursion Relations for Effective Field Theories}},
  \href{https://doi.org/10.1103/PhysRevLett.116.041601}{\emph{Phys. Rev. Lett.}
  {\bfseries 116} (2016) 041601}
  [\href{https://arxiv.org/abs/1509.03309}{{\ttfamily 1509.03309}}].

\bibitem{Bern2019a}
Z.~Bern, J.J.~Carrasco, M.~Chiodaroli, H.~Johansson and R.~Roiban, \emph{{The
  Duality Between Color and Kinematics and its Applications}},
  \href{https://arxiv.org/abs/1909.01358}{{\ttfamily 1909.01358}}.

\bibitem{Bern2022}
Z.~Bern, J.J.~Carrasco, M.~Chiodaroli, H.~Johansson and R.~Roiban, \emph{{The
  SAGEX review on scattering amplitudes Chapter 2: An invitation to
  color-kinematics duality and the double copy}},
  \href{https://doi.org/10.1088/1751-8121/ac93cf}{\emph{J. Phys.} {\bfseries
  A55} (2022) 443003} [\href{https://arxiv.org/abs/2203.13013}{{\ttfamily
  2203.13013}}].

\bibitem{Adamo2022}
T.~Adamo, J.J.M.~Carrasco, M.~Carrillo-Gonz\'alez, M.~Chiodaroli, H.~Elvang,
  H.~Johansson et~al., \emph{{Snowmass White Paper: the Double Copy and its
  Applications}},  in \emph{{Snowmass 2021}}, 4, 2022
  [\href{https://arxiv.org/abs/2204.06547}{{\ttfamily 2204.06547}}].

\bibitem{Bern:2010ue}
Z.~Bern, J.J.M.~Carrasco and H.~Johansson, \emph{{Perturbative Quantum Gravity
  as a Double Copy of Gauge Theory}},
  \href{https://doi.org/10.1103/PhysRevLett.105.061602}{\emph{Phys. Rev. Lett.}
  {\bfseries 105} (2010) 061602}
  [\href{https://arxiv.org/abs/1004.0476}{{\ttfamily 1004.0476}}].

\bibitem{Cachazo:2013gna}
F.~Cachazo, S.~He and E.Y.~Yuan, \emph{{Scattering equations and
  Kawai-Lewellen-Tye orthogonality}},
  \href{https://doi.org/10.1103/PhysRevD.90.065001}{\emph{Phys. Rev.}
  {\bfseries D90} (2014) 065001}
  [\href{https://arxiv.org/abs/1306.6575}{{\ttfamily 1306.6575}}].

\bibitem{Cachazo:2013iea}
F.~Cachazo, S.~He and E.Y.~Yuan, \emph{{Scattering of Massless Particles:
  Scalars, Gluons and Gravitons}},
  \href{https://doi.org/10.1007/JHEP07(2014)033}{\emph{JHEP} {\bfseries 07}
  (2014) 033} [\href{https://arxiv.org/abs/1309.0885}{{\ttfamily 1309.0885}}].

\bibitem{Cachazo2014a}
F.~Cachazo, S.~He and E.Y.~Yuan, \emph{{Scattering of Massless Particles in
  Arbitrary Dimensions}},
  \href{https://doi.org/10.1103/PhysRevLett.113.171601}{\emph{Phys. Rev. Lett.}
  {\bfseries 113} (2014) 171601}
  [\href{https://arxiv.org/abs/1307.2199}{{\ttfamily 1307.2199}}].

\bibitem{Monteiro2011}
R.~Monteiro and D.~O'Connell, \emph{{The Kinematic Algebra From the Self-Dual
  Sector}}, \href{https://doi.org/10.1007/JHEP07(2011)007}{\emph{JHEP}
  {\bfseries 07} (2011) 007} [\href{https://arxiv.org/abs/1105.2565}{{\ttfamily
  1105.2565}}].

\bibitem{Cheung2021}
C.~Cheung and J.~Mangan, \emph{{Covariant color-kinematics duality}},
  \href{https://doi.org/10.1007/JHEP11(2021)069}{\emph{JHEP} {\bfseries 11}
  (2021) 069} [\href{https://arxiv.org/abs/2108.02276}{{\ttfamily
  2108.02276}}].

\bibitem{Arkani-Hamed:2023mvg}
N.~Arkani-Hamed, H.~Frost, G.~Salvatori, P.-G.~Plamondon and H.~Thomas,
  \emph{{All Loop Scattering For All Multiplicity}},
  \href{https://arxiv.org/abs/2311.09284}{{\ttfamily 2311.09284}}.

\bibitem{ArkaniHamed2024}
N.~Arkani-Hamed, Q.~Cao, J.~Dong, C.~Figueiredo and S.~He, \emph{{Nonlinear
  Sigma model amplitudes to all loop orders are contained in the $Tr(\Phi^3)$
  theory}}, \href{https://doi.org/10.1103/PhysRevD.110.065018}{\emph{Phys.
  Rev.} {\bfseries D110} (2024) 065018}
  [\href{https://arxiv.org/abs/2401.05483}{{\ttfamily 2401.05483}}].

\bibitem{Kampf2013b}
K.~Kampf, J.~{Novotn\'y} and J.~Trnka, \emph{{Tree-level amplitudes in the
  nonlinear sigma model}},
  \href{https://doi.org/10.1007/JHEP05(2013)032}{\emph{JHEP} {\bfseries 05}
  (2013) 032} [\href{https://arxiv.org/abs/1304.3048}{{\ttfamily 1304.3048}}].

\bibitem{Bartsch2024a}
C.~Bartsch, K.~Kampf, J.~Novotny and J.~Trnka, \emph{{All-loop soft theorem for
  pions}}, \href{https://doi.org/10.1103/PhysRevD.110.045009}{\emph{Phys. Rev.}
  {\bfseries D110} (2024) 045009}
  [\href{https://arxiv.org/abs/2401.04731}{{\ttfamily 2401.04731}}].

\bibitem{Arkani-Hamed:2024yvu}
N.~Arkani-Hamed and C.~Figueiredo, \emph{{Circles and Triangles, the NLSM and
  Tr($\Phi^3$)}},  \href{https://arxiv.org/abs/2403.04826}{{\ttfamily
  2403.04826}}.

\bibitem{Cheung2015a}
C.~Cheung, K.~Kampf, J.~{Novotn\'y} and J.~Trnka, \emph{{Effective Field
  Theories from Soft Limits of Scattering Amplitudes}},
  \href{https://doi.org/10.1103/PhysRevLett.114.221602}{\emph{Phys. Rev. Lett.}
  {\bfseries 114} (2015) 221602}
  [\href{https://arxiv.org/abs/1412.4095}{{\ttfamily 1412.4095}}].

\bibitem{Cheung2017a}
C.~Cheung, K.~Kampf, J.~Novotn{\'y}, C.-H.~Shen and J.~Trnka, \emph{{A periodic
  table of effective field theories}},
  \href{https://doi.org/10.1007/JHEP02(2017)020}{\emph{JHEP} {\bfseries 02}
  (2017) 020} [\href{https://arxiv.org/abs/1611.03137}{{\ttfamily
  1611.03137}}].

\bibitem{Arkani-Hamed:2023lbd}
N.~Arkani-Hamed, H.~Frost, G.~Salvatori, P.-G.~Plamondon and H.~Thomas,
  \emph{{All Loop Scattering As A Counting Problem}},
  \href{https://arxiv.org/abs/2309.15913}{{\ttfamily 2309.15913}}.

\bibitem{Neeling2022}
D.~de~Neeling, D.~Roest and S.~Veldmeijer, \emph{{Flavour-kinematics duality
  for Goldstone modes}},
  \href{https://doi.org/10.1007/JHEP10(2022)066}{\emph{JHEP} {\bfseries 10}
  (2022) 066} [\href{https://arxiv.org/abs/2204.11629}{{\ttfamily
  2204.11629}}].

\bibitem{Cachazo2015}
F.~Cachazo, S.~He and E.Y.~Yuan, \emph{{Scattering Equations and Matrices: From
  Einstein To Yang-Mills, DBI and NLSM}},
  \href{https://doi.org/10.1007/JHEP07(2015)149}{\emph{JHEP} {\bfseries 07}
  (2015) 149} [\href{https://arxiv.org/abs/1412.3479}{{\ttfamily 1412.3479}}].

\bibitem{Boulware1968}
D.G.~Boulware and L.S.~Brown, \emph{{Tree Graphs and Classical Fields}},
  \href{https://doi.org/10.1103/PhysRev.172.1628}{\emph{Phys. Rev.} {\bfseries
  172} (1968) 1628}.

\bibitem{Brauner2024}
T.~Brauner, \emph{{Effective Field Theory for Spontaneously Broken Symmetry}},
  vol.~1023 of \emph{Lecture Notes in Physics}, Springer (2024),
  \href{https://doi.org/10.1007/978-3-031-48378-3}{10.1007/978-3-031-48378-3},
  [\href{https://arxiv.org/abs/2404.14518}{{\ttfamily 2404.14518}}].

\bibitem{Bando1988a}
M.~Bando, T.~Kugo and K.~Yamawaki, \emph{{Nonlinear Realization and Hidden
  Local Symmetries}},
  \href{https://doi.org/10.1016/0370-1573(88)90019-1}{\emph{Phys. Rept.}
  {\bfseries 164} (1988) 217}.

\bibitem{Li:2024bwq}
Y.~Li, T.~Wang, T.~Brauner and D.~Roest, \emph{{Diagrammatic Derivation of
  Hidden Zeros and Exact Factorisation of Pion Scattering Amplitudes}},
  \href{https://arxiv.org/abs/2412.14858}{{\ttfamily 2412.14858}}.

\bibitem{Bogers2018b}
M.P.~Bogers and T.~Brauner, \emph{{Lie-algebraic classification of effective
  theories with enhanced soft limits}},
  \href{https://doi.org/10.1007/JHEP05(2018)076}{\emph{JHEP} {\bfseries 05}
  (2018) 076} [\href{https://arxiv.org/abs/1803.05359}{{\ttfamily
  1803.05359}}].

\bibitem{Adams:2006sv}
A.~Adams, N.~Arkani-Hamed, S.~Dubovsky, A.~Nicolis and R.~Rattazzi,
  \emph{{Causality, analyticity and an IR obstruction to UV completion}},
  \href{https://doi.org/10.1088/1126-6708/2006/10/014}{\emph{JHEP} {\bfseries
  10} (2006) 014} [\href{https://arxiv.org/abs/hep-th/0602178}{{\ttfamily
  hep-th/0602178}}].

\bibitem{Volkov1973a}
D.V.~Volkov, \emph{{Phenomenological Lagrangians}}, {\emph{Sov. J. Part. Nucl.}
  {\bfseries 4} (1973) 1}.

\bibitem{Ogievetsky1974a}
V.I.~Ogievetsky, \emph{{Nonlinear realizations of internal and spacetime
  symmetries}}, {\emph{Acta Univ. Wratislav.} {\bfseries 207} (1974) 117}.

\bibitem{Ivanov1975a}
E.~Ivanov and V.I.~Ogievetsky, \emph{{Inverse Higgs effect in nonlinear
  realizations}}, {\emph{Theor. Math. Phys.} {\bfseries 25} (1975) 1050}.

\bibitem{Hinterbichler2015a}
K.~Hinterbichler and A.~Joyce, \emph{{Hidden symmetry of the Galileon}},
  \href{https://doi.org/10.1103/PhysRevD.92.023503}{\emph{Phys. Rev.}
  {\bfseries D92} (2015) 023503}
  [\href{https://arxiv.org/abs/1501.07600}{{\ttfamily 1501.07600}}].

\bibitem{Roest2021}
D.~Roest, \emph{{The Special Galileon as Goldstone of Diffeomorphisms}},
  \href{https://doi.org/10.1007/JHEP01(2021)096}{\emph{JHEP} {\bfseries 01}
  (2021) 096} [\href{https://arxiv.org/abs/2004.09559}{{\ttfamily
  2004.09559}}].

\bibitem{Novotny2017a}
J.~Novotn{\'y}, \emph{{Geometry of special Galileons}},
  \href{https://doi.org/10.1103/PhysRevD.95.065019}{\emph{Phys. Rev.}
  {\bfseries D95} (2017) 065019}
  [\href{https://arxiv.org/abs/1612.01738}{{\ttfamily 1612.01738}}].

\bibitem{GarciaSaenz2019}
S.~Garcia-Saenz, J.~Kang and R.~Penco, \emph{{Gauged Galileons}},
  \href{https://doi.org/10.1007/JHEP07(2019)081}{\emph{JHEP} {\bfseries 07}
  (2019) 081} [\href{https://arxiv.org/abs/1905.05190}{{\ttfamily
  1905.05190}}].

\bibitem{Preucil2019}
F.~P\v{r}eu\v{c}il and J.~Novotn\'y, \emph{{Special Galileon at one loop}},
  \href{https://doi.org/10.1007/JHEP11(2019)166}{\emph{JHEP} {\bfseries 11}
  (2019) 166} [\href{https://arxiv.org/abs/1909.06214}{{\ttfamily
  1909.06214}}].

\bibitem{ArkaniHamed2023}
N.~Arkani-Hamed, Q.~Cao, J.~Dong, C.~Figueiredo and S.~He, \emph{{Hidden zeros
  for particle/string amplitudes and the unity of colored scalars, pions and
  gluons}}, \href{https://doi.org/10.1007/JHEP10(2024)231}{\emph{JHEP}
  {\bfseries 10} (2024) 231}
  [\href{https://arxiv.org/abs/2312.16282}{{\ttfamily 2312.16282}}].

\bibitem{Brauner2021a}
T.~Brauner, \emph{{Exceptional nonrelativistic effective field theories with
  enhanced symmetries}},
  \href{https://doi.org/10.1007/JHEP02(2021)218}{\emph{JHEP} {\bfseries 02}
  (2021) 218} [\href{https://arxiv.org/abs/2008.12078}{{\ttfamily
  2008.12078}}].

\bibitem{Mojahed2022}
M.A.~Mojahed and T.~Brauner, \emph{{Nonrelativistic effective field theories
  with enhanced symmetries and soft behavior}},
  \href{https://doi.org/10.1007/JHEP03(2022)086}{\emph{JHEP} {\bfseries 03}
  (2022) 086} [\href{https://arxiv.org/abs/2201.01393}{{\ttfamily
  2201.01393}}].

\bibitem{Cheung2022}
C.~Cheung, A.~Helset and J.~Parra-Martinez, \emph{{Geometric soft theorems}},
  \href{https://doi.org/10.1007/JHEP04(2022)011}{\emph{JHEP} {\bfseries 04}
  (2022) 011} [\href{https://arxiv.org/abs/2111.03045}{{\ttfamily
  2111.03045}}].

\bibitem{Cheung2022c}
C.~Cheung, A.~Helset and J.~Parra-Martinez, \emph{{Geometry-kinematics
  duality}}, \href{https://doi.org/10.1103/PhysRevD.106.045016}{\emph{Phys.
  Rev.} {\bfseries D106} (2022) 045016}
  [\href{https://arxiv.org/abs/2202.06972}{{\ttfamily 2202.06972}}].

\bibitem{Craig2023a}
N.~Craig and Y.-T.~Lee, \emph{{Effective Field Theories on the Jet Bundle}},
  \href{https://doi.org/10.1103/PhysRevLett.132.061602}{\emph{Phys. Rev. Lett.}
  {\bfseries 132} (2024) 061602}
  [\href{https://arxiv.org/abs/2307.15742}{{\ttfamily 2307.15742}}].

\bibitem{Helset2024}
A.~Helset, \emph{{Color-kinematics duality for nonlinear sigma models with
  nonsymmetric cosets}},
  \href{https://doi.org/10.1103/PhysRevD.110.L101701}{\emph{Phys. Rev.}
  {\bfseries D110} (2024) L101701}
  [\href{https://arxiv.org/abs/2406.10955}{{\ttfamily 2406.10955}}].

\end{thebibliography}\endgroup

\end{fmffile}

\end{document}